\documentclass[sn-mathphys-num]{sn-jnl}

\usepackage{aas_macros}

\usepackage{graphicx}%
\usepackage{multirow}%
\usepackage{amsmath,amssymb,amsfonts}%
\usepackage{amsthm}%
\usepackage{mathrsfs}%
\usepackage[title]{appendix}%
\usepackage{xcolor}%
\usepackage{textcomp}%
\usepackage{manyfoot}%
\usepackage{booktabs}%
\usepackage{algorithm}%
\usepackage{algorithmicx}%
\usepackage{algpseudocode}%
\usepackage{listings}%
\usepackage{ulem}%
\theoremstyle{thmstyleone}%
\theoremstyle{thmstyletwo}%

\theoremstyle{thmstylethree}%

\graphicspath{{./Figs_PDF/}}

\begin{document}
\title{Large fluctuations and cusp-core transformation in dwarf galaxies}


\author[1]{\fnm{Mahmoud} \sur{Hashim}}\email{mahmoud.hashim@bue.edu.eg}

\author[1]{\fnm{Amr A.} \sur{El-Zant}}\email{amr.elzant@bue.edu.eg}

\author[2,3]{\fnm{Fran\c{c}oise} \sur{Combes}}\email{francoise.combes@obspm.fr}

\affil[1]{\orgdiv{Centre for Theoretical Physics}, \orgname{The British University in Egypt}, \orgaddress{\street{Sherouk City}, \city{Cairo}, \postcode{11837}, \country{Egypt}}}

\affil[2]{\orgdiv{LUX}, \orgname{Observatoire de Paris}, \orgaddress{\street{CNRS, PSL Univ., Sorbonne Univ.}, \city{Paris}, \postcode{F-75014}, \country{France}}}

\affil[3]{\orgdiv{Collège de France}, \orgaddress{\street{11 place Marcelin Berthelot}, \city{Paris}, \postcode{F-75005}, \country{France}}}


\abstract{

Dwarf galaxies are abundant and dark matter dominated, rendering them particularly relevant for testing a cosmological model dominated by a dark sector. Whether their density profiles agree with $\Lambda$CDM expectations has long been a point of contention, dubbed the cusp/core problem. While numerical simulations have suggested that the erasure of central dark matter cusps may arise from potential fluctuations in feedback-driven gas, whether the required fluctuation levels are achieved in actual galaxies has remained uncertain. 
By measuring the power spectra of neutral gas density contrasts in a sample of gas-rich dwarf galaxies, we show that the observed fluctuations match theoretical expectations when their impact is derived from a stochastic dynamical model. Specifically, the relaxation times for core formation at the observationally estimated radii are generally on the order of a Hubble time or less. Core sizes are therefore consistent with conservative estimates assuming their generation through steady fluctuations acting over cosmic history. The relaxation time increases with halo mass, suggesting that core formation via this mechanism becomes less efficient in more massive systems.}


%



\maketitle

\section*{Main}

Astronomical observations,  interpreted in terms of general 
relativity and Newtonian dynamics, 
imply that our universe is dominated 
by a dark sector. In the standard $\Lambda$CDM scenario, this consists
of dark energy, represented by a cosmological constant $\Lambda$, and  cold dark matter 
(CDM)~\cite{LidLah2026}.  
Phenomenologically, the dark energy is particularly simple; 
its density is homogeneous in space and time. 
The description of the dynamics of the 
dark matter is more involved, as it can collapse into nonlinear haloes with 
average overdensities of the order of a hundred relative to the 
cosmological background density. Nevertheless, the structure of such haloes
is now well established through numerical simulations; in particular, it has  long been known that their central density diverges as one zooms in towards the centre down to 
the resolution limit~\cite{DBC91}. 
However, soon after the results of the first simulations displaying 
central density cusps became available, it was pointed out that the dynamics 
of dark matter dominated galaxies, as inferred from observations, may imply less steep 
central density
profiles~\cite{1994Natur.370..629M}. 
Indeed, although there is a diversity in  galaxy 
 rotation curves, perhaps partly born of systematic effects 
 related to non-circular motions, at least some dwarfs appear 
 to display non-singular central density profiles, and 
 generally contain less dark matter in the central region than is expected from dissipationless cosmological 
 simulations~\cite{2015PNAS..11212249W, BullockRev2017}.  

In other words, the dark matter structures invoked in part to explain the outer dynamics of galaxies may not fit those of the inner parts. 
Solutions to this longstanding 'cusp/core' problem have invoked variations on the dark matter model, in attempts to dynamically `heat'  the centres of haloes and decrease their concentrations.  These include pre-heated warm dark matter~\cite{2000ApJ...542..622C}, thermally conducting self-interacting dark matter~\cite{2000PhRvL..84.3760S}, and  zero-point quantum fluctuations in fuzzy dark matter~\cite{2000PhRvL..85.1158H}. 
But cold dark matter can itself be heated by  
gaining random energy from massive gaseous or stellar clumps via dynamical  
friction~\cite{EZSH2001}, or through sudden changes in the potential triggered by rapid feedback (e.g. starburst) driven gas causing  outflows~\cite{1996MNRAS.283L..72N}. 
In the latter case, numerous outflows and inflows,  from and towards 
the central region, are most effective in transforming the density cusp into a core~\cite{2024arXiv241209566M}. 
The resulting irreversible dynamics  
resemble those  stemming from massive clumps imparting random energy 
to the dark matter as they move through it~\cite{Mash2006Natur, PG2012MNRAS, PGNatur2014, EZFC, HEZFRC}.  
Observationally, however, it may be possible to distinguish between the
two mechanisms, through star formation histories and the associated 
energetics of the feedback that drives large scale gas motions.  
Thus, while both mechanisms may act in tandem~\cite{2021MNRAS.504.3509O}, with 
dynamical friction from massive clumps being efficient 
in large galaxies, simultaneously 
forming halo cores and stellar bulges~\cite{HashimEZPop2024}, feedback-driven fluctuations 
may be the dominant mechanism  in dwarfs~\cite{2014MNRAS.437..415D,2019MNRAS.484.1401R}. 
But modelling the physical processes that may lead to 
the cusp/core transformation in this context is 
far more involved than simulating the dark matter, and the conclusions 
are consequently much less robust or secure.  

In standard cosmology, galaxies form as gas settles into the centres of CDM haloes, becoming dynamically important and forming stars~\cite{WR1978}. 
Beyond an early period of weak force mediated thermal equilibrium, which sets the scale of the smallest objects, the formation of CDM structures may be  described in terms of Newtonian gravitational dynamics on scales significantly smaller than the cosmological horizon, relevant to galaxies. The baryons, on the other hand, can also interact electromagnetically and through the strong nuclear force. They are thus 
subject to complex hydrodynamic forces and thermal and feedback processes in a highly inhomogeneous multiphase medium. These may be modelled through large scale numerical simulations, but subgrid models are required to describe star formation and stellar/AGN feedback, and their mode of energy injection and transfer~\cite{Vogelsberger2020NatRP}. 

Such simulations generally reproduce the observed 
low baryon to dark matter content in dwarf galaxies, resulting from a shallow potential well. This property renders dwarf galaxies  
dark matter dominated even in the central region, and consequently makes for ideal  
laboratories for testing predicted dark matter density profiles
against observations~\cite{2013MNRAS.429.3068T, 2016MNRAS.462.3628R, 2019MNRAS.482..821O, Cooke2022MNRAS, 2023MNRAS.521.1316R, deLeo2024MNRAS,
 2024MNRAS.529.2379R}.  
However, whether cusps are transformed into cores in a given hydrodynamic simulation depends on the details of 
the implemented subgrid modelling. 
It has been long suspected that the key ingredient involved is the star formation threshold---with a high threshold
leading to conditions favourable for core formation and a low
threshold being unfavourable~\cite{PG2012MNRAS, PGNatur2014, 2019MNRAS.488.2387B, 2020MNRAS.499.2648D, 2019MNRAS.486.4790B}. But it has recently been realised that additional details of the modelling of the subgrid physics of feedback and energy 
transfer may also be important~\cite{2023MNRAS.520..461J}. 

The key dynamical condition for core formation in CDM-based simulations with feedback appears to be the presence of sustained fluctuations in the gravitational potential of the gas, which couple to the dynamics of the dark matter. Ref.~\cite{EZFC} proposed that such a situation is amenable to analysis in terms of first dynamical 
principles if the fluctuations form a stochastic process 
described by a power spectrum. 
In this case, the dynamical 
`heating' of the dark matter particles can be given a precise meaning, 
through the random energy they gain from stochastic 
fluctuations.
The effect is predicted to  depend primarily on two parameters: the gas density and the amplitude of the spectrum of density fluctuations. It thereby bypasses the complex details of the baryonic physics of galaxy formation.
The assumptions and consequences of this theoretical model were tested against the results of a numerical simulation of a model dwarf galaxy in ref.~\cite{HEZFRC}.

Although gravitational potential fluctuations in the gas cannot be observed,  the  model  
maps these to the density field and determines the dynamical response of halo particles. It thus 
provides a pathway for directly 
testing whether the dynamics driven by these 
may indeed lead to the cusp-core transformation.
In this context, observations of local galaxies should 
provide a lower limit of the most important  parameters, 
as the gas density should on average be smaller than  at earlier 
times (due to  outflows, and star formation), and likewise with the amplitude of the fluctuations  (as feedback is presumably stronger at earlier times). Measuring these quantities in local galaxies should thus provide an upper limit of the relaxation time for core formation. 
And with this comes a  direct test of the proposal of cusp/core
transformation through potential fluctuations, 
independently of the 
complexities involved in numerical and subgrid implementations inherent to  simulations of the process. This is what we set out to do here.

\section*{Power spectra}

The bulk of the gas mass in dwarf irregular 
galaxies is in atomic form~\cite{2024ARA&A..62..113H}. 
In this medium, density fluctuations arise from the presence of gas clouds, 
feedback-driven voids, shells and turbulent structures. 
We choose a sample of such galaxies from the LITTLE THINGS HI galaxy survey of local ($< 10 \rm\; Mpc$) dwarfs, which has sufficient resolution 
(110 pc at mean distance 3.7 Mpc), 
to capture the aforementioned structures~\cite{2012AJ144134H}. 
The survey includes 
gas-rich galaxies with absolute visual magnitudes between 
-9 and -18. They are drawn from earlier databases where 
morphological inclusion criteria were also used~\cite{2006ApJS..162...49H,2004AJ....128.2170H}. 
The total sample includes 37 dwarf irregulars 
and 4 blue compact dwarfs.

\begin{figure}
    \centering
    \includegraphics[width=\linewidth]{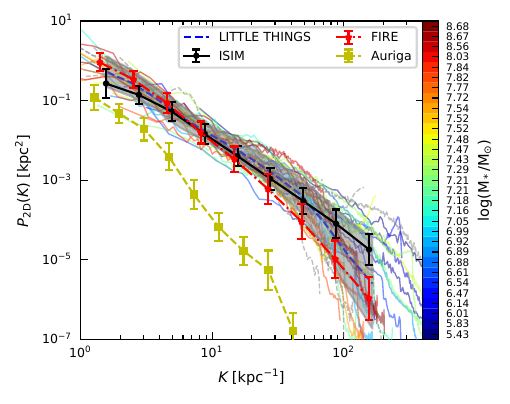}
    \caption{Collected power spectra of  HI surface density fluctuations $\delta \Sigma /\Sigma$, in LITTLE THINGS galaxies with stellar mass denoted on the right, and shaded regions showing the dispersion. The solid line corresponds to the gas fluctuation spectrum in the hydrodynamic simulation of a dwarf galaxy~\cite{2016MNRAS.462.3628R, HEZFRC}, referred to as ISIM, used to study the cusp-core transformation in ref.~\cite{HEZFRC} (the mean and dispersion are over snapshots between  4 Gyr and a Hubble time). The spectra from observations are inferred from images generated from the data (as described in Sections~\ref{sec:sudens} and~\ref{sec:obsps}). The largest wavenumbers correspond to the pixel sizes of the images
    (spectra with beam size resolution are shown in Fig.~\ref{fig:Pixel_obs});  the smallest wavenumbers are determined by the FFT box size, which is double the radius containing $90 \%$ of the observed HI gas mass. For comparison, we show an example power spectrum from the Auriga simulation (level 3/dwarf, ID~1~\cite{AURIGA2024}),  where no halos cores were found to form~\cite{2019MNRAS.486.4790B}.      
    Here, a low star formation threshold is regulated by a long star formation timescale and stiff equation of state, suppressing density fluctuations.  Also shown is a spectrum of gas fluctuations  in a halo from the FIRE-2 cosmological simulation where a core does form (identified as $\rm m11d$ in~\cite{FIRE-2-Pub2023ApJS}).}
\label{fig:collect_Spectra}
\end{figure}

We produced power spectra of relative  surface 
density fluctuations
$\delta \Sigma/\Sigma = \Sigma({\bf R})/\Sigma(R) - 1$
 (as detailed in Section~\ref{sec:obsps}), 
 where  $\Sigma({\bf R})$ is the surface density of the gas at position ${\bf R}$  on the image of the galaxy and $\Sigma (R)$ refers to averages over rings at radius 
$\vert {\bf R} \vert$. This is a measure of the level of density perturbations
that is independent of the absolute value of the average density.  
In addition to being theoretically relevant (as we will see below), it 
is a dimensionless measure facilitating comparison between different systems.  

The resulting power spectra for the whole sample of galaxies are
shown in Fig.~\ref{fig:collect_Spectra}, 
where we also plot the time average 
of the corresponding  spectra in the simulation analysed
in ref.~\cite{HEZFRC} (referred to as ISIM), 
which employs a feedback recipe with 
high star formation threshold (300 atoms ${\rm cm^{-3}}$ and 
and efficiency  $\epsilon_{\rm eff}= 0.1$ per local free fall time, 
allowing for 
bursty episodic feedback. The gas was simulated with 
maximum resolution 
of about 4~pc,  to capture momentum driving 
from individual supernovae events. 
The recipe led to results that 
 compared well with 
various  properties of observed 
dwarf irregular galaxies, including rotation curves, 
stellar surface density, metallicity, gas and stellar 
velocity dispersion and star formation 
history~\cite{2016MNRAS.459.2573R, 2016MNRAS.462.3628R}, 
despite the isolated setting of the simulation. 
On the other hand, this property renders the results  
ideal for testing the theoretical model, in 
order to capture and isolate the central physical and dynamical 
features of the core formation process in a controlled setting. 

 We have also examined the power spectra of gas embedded in simulated cosmological haloes, as illustrated in Fig.~\ref{fig:collect_Spectra}. The FIRE-2 cosmological simulation~\cite{FIRE-2-Pub2023ApJS} employs a feedback recipe with a high star formation threshold, broadly similar to ISIM, 
which allows for core formation~~\cite{2015MNRAS.454.2092O}. 
For haloes that formed cores in FIRE, the resulting power spectra were likewise consistent with the data within the uncertainties. In contrast, results for the gas embedded in haloes 
from the Auriga simulation~\cite{AURIGA2024}, were quite different. There,  a low star formation threshold is adopted, and feedback is regulated by a long star formation timescale and a stiff equation of state. This results in less episodic star formation and, crucially, in the absence of large density fluctuations. Consequently, the spectra display substantial quantitative and qualitative differences, reflected in their normalization and tilt, respectively, as can be readily seen from the example spectrum included in Fig.~\ref{fig:collect_Spectra}. We present further details regarding the analysis of these cosmological simulations in Section~\ref{sec:cosmo}.

In ref~\cite{HEZFRC} 
it was shown that the gas density fluctuations
in ISIM, as reflected in the power spectrum, can
explain the cusp/core inferred transformation in the context of our model.   
Fig.~\ref{fig:collect_Spectra} shows that  the normalizations and slopes of the spectra 
of the observed galaxies are distributed around the line corresponding to the simulation. 
This appears remarkable, given that the average 
in the simulation is taken with a starting time of 4 Gyr, when a significant 
core starts forming, to a Hubble time, while
the power spectra inferred from the observed sample correspond to  local galaxies. (Note that these are the measured 
spectra in physical units, with no scaling introduced.) This may suggest that the density and potential fluctuations  may indeed 
be  approximately represented  by quasi-stationary  
processes, as is assumed in the model and approximately realized in the simulation. 

As the density profiles of LITTLE THINGS galaxies  are generally consistent with  
having halo cores rather than cusps (Section~\ref{sec:slopes}), 
the matching of their gas spectra with those of gas  embedded in core-forming haloes (in ISIM and FIRE)---in contrast to the case of Auriga, where no cores were found to form~\cite{2019MNRAS.486.4790B}---naturally leads to the following question: have the fluctuations characterized by the 
inferred spectra led to a cusp/core transformation in the observed galaxies, 
in accordance with expectations from our model?
We test for this in the coming section. 

\section*{The relaxation time of core formation}

The physical picture underlying our model  is as follows. 
As the gas settles into the centre of a dark matter halo,
a critical density threshold for star formation is reached. 
If that threshold is sufficiently large, star formation 
leads to a prolonged episodic feedback process, accompanied 
by repeated large scale outflows and inflows 
confined within the central region of the dark matter halo
(as opposed to escaping the potential well). 
Within that region, the large scale flows carry  smaller scale 
density fluctuations, which 
may be represented by 
a three dimensional (3D)
power spectrum $P_{3D} (k) \equiv P (k)$.   
In analogy with compressible turbulence and observations 
of the interstellar medium, reference~\cite{EZFC} 
assumed that 
$P(k)$ is a power law between 
maximal and minimal wavenumbers 
$k_x$ and $k_m$ respectively. 
As we saw, a power law within a certain 
interval is also an adequate description 
of the two dimensional (2D) power spectra of 
the HI gas in LITTLE THINGS galaxies. 

A summary of the main features of the 
dynamical model developed in~\cite{EZFC} 
is given in Section~\ref{sec:model}.
In its context, gaseous density perturbations
are readily transformed into potential fluctuations through the Poisson equation. 
The  resulting fluctuating force imparts random velocities to the  
dark matter particles moving through the gas, 
into higher energy orbits away from the centre. This 
reduces the central density, transforming the cusp into a core.
The process takes place over a relaxation time, 
over which the random velocities gained by the particles become 
of the order of their orbital speed. This timescale 
is given by
\begin{equation}
    t_{\rm relax} = \frac{n v_p^2 v_r}{8\pi (G\rho_0)^2 P (k_m)}. 
\label{eq:relax}
\end{equation}
Here, $n > 0$ is the index of the power law spectrum, $P(k) \propto k^{-n}$;
$P(k_m)$ is the value of the 3D power 
spectrum at the minimal wavenumber $k_m$;  
$\rho_0$ is the characteristic 
3D gas density, evaluated 
at the projected half-mass 
radius (Section~\ref{sec:rho0});
$v_p$ is the typical velocity of a dark matter 
particle affected by the stochastic fluctuation field; 
and $v_r$ is the velocity of the large scale gaseous flows
carrying the density fluctuations. 
This equation is a generalization of Chandrasekhar's  
two body relaxation 
formula when $n \ne 0$.  
The latter corresponds  to
a white noise power spectrum, 
$P = 1/N$ (with $N$ the particle number density)
and Coulomb logarithm $\Lambda = k_x/k_m$~\cite{EZFCH}.  
For $n \ne 0$, equation~(\ref{eq:relax}) 
replaces the usual formula, with  $n/P(k_m)$  
standing for $1/\left(P \ln \Lambda\right)$.
It is customary in applying  Chandrasekhar's formula 
to realistic (inhomogeneous) systems, 
to assume a local density, possibly averaged within a certain 
radius (e.g., the half mass radius), while the maximum impact parameter
is assumed to correspond to the entire size of the system. 
In the present case,  
while the characteristic density is defined in terms of the half mass radius, 
the spectra are evaluated at the $90 \%$ radius,  
in order to capture the power at the largest scales   
associated with $k_m$ and ensure convergence of the spectra (Section~\ref{sec:obsps}).
 Note that the dependence of $t_{\rm relax}$ on the largest fluctuations---when the spectrum displays a well-defined power-law region between $k_m$ and $k_x$---renders its estimation robust against small-scale noise and limited resolution.

As the index $n$ and normalization $P(k_m)$ refer to 3D spectra, to connect with observations we calibrate the 2D spectra into 3D ones using simulations and a projection model (as described in Section~\ref{sec:simps}).
We find that, for the parameters of the galaxies at hand, 
the index $n_{3D} \equiv n \simeq 1.1~n_{2D}$.  
The maximal fluctuation scales (corresponding to $k_m$) are set through the maximal scales  
of HI cloud structures, estimated for each galaxy 
(Section~\ref{sec:obsps}).  
This has the advantage 
of being independent from the transformation of the 
power spectrum from 2D to 3D and insensitive 
to large scale deformations in the gas patterns (e.g., spiral structures).

For a subset of the galaxies in LITTLE THINGS, 
high resolution mass models are available~\cite{2015AJ....149..180O}, 
including fits to the halo mass distributions  
using isothermal profiles with quasi-constant density inside 
core radius $r_c$. 
We use these to estimate  
the characteristic velocities entering into equation~(\ref{eq:relax}),  
by  fixing  $v_p$ to the circular velocity at $r_c$. In this way 
$t_{\rm relax} (r_c)$ estimates the timescale over which 
a core of the observed size $r_c$ can be formed through 
the effect of potential fluctuations. 
The assumption here is that the core radius reflects the scale inside which an initial inner density cusp is significantly altered. For, although the existence of a precisely zero inner logarithmic slope for the density profile is generally difficult to confirm observationally (Section~\ref{sec:slopes}), for most galaxies in the sample the cored profile was found to be strongly 
favoured by the data, while a steep inner cusp is entirely inconsistent with it~\cite{2015AJ....149..180O}.
Note that because the circular speed increases with radius (as does the velocity dispersion in an initial cusp), the relaxation time satisfies \(t_{\rm relax}(r \le r_c) \le t_{\rm relax}(r_c)\). Consequently, if gas fluctuations significantly affect particles at \(r_{c}\), they will also affect all particles within \(r_{c}\). The criterion is  tested in Fig.~\ref{fig:core_relax}.

Following refs.~\cite{EZFC} and \cite{HEZFRC}, we estimate the velocity of large-scale gas flows—assumed to carry the density fluctuations—as \(v_r = (d/2)/t_D(d/2)\), where \(t_D(d/2) = 1/\sqrt{G \bar{\rho}_{\rm DM}(d/2)}\) (with \(\bar{\rho}_{\rm DM}(d/2)\) being the average dark matter density inside \(d/2\)). This form of \(v_{r}\) implies a characteristic diameter \(d\) within which feedback-driven fountain flows are contained, maintaining a nearly constant total gas mass inside \(d/2\) even as the inner mass fluctuates strongly (cf.\ Fig.~2 of ref.~\cite{HEZFRC}). Because feedback-driven fluctuations diminish rapidly beyond \(d/2\) by definition, previous studies assumed this boundary to be the scale at which the power spectra are evaluated~\cite{EZFC, HEZFRC}. We adopt the same assumption here; for our galaxy sample, we equate \(d/2\) to the convergence scale of the power spectra, \(R_{90}\). Evaluating \(v_{r}\) at \(R_{90}\) is conservative in the sense that \(R_{90}\) typically lies in the flat region of the rotation curve, making \(v_{r}\) insensitive to exact radial positioning. Conversely, evaluating it at smaller radii yields shorter relaxation times \(t_{\mathrm{r}elax}\) (as noted above in the case of \(v_{p}\)).

\begin{figure}
    \centering
    \includegraphics[width= \textwidth]{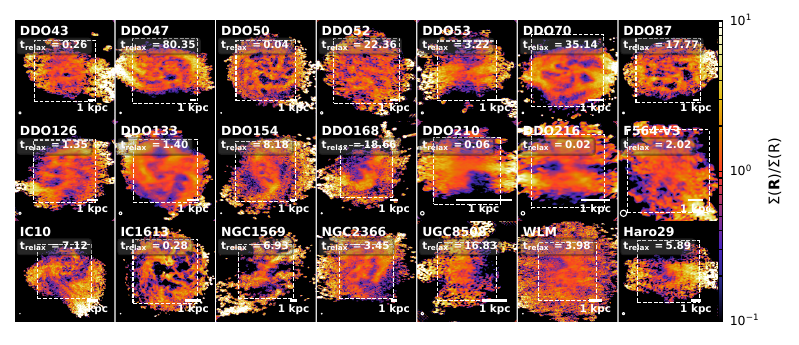}
    \caption{Maps of relative surface density 
    $\Sigma ({\bf R})/\Sigma (R)$.  
    The quoted times correspond to the timescales (in Gyr) required (according to 
    equation~\ref{eq:relax}) to produce halo density cores with sizes consistent with those obtained from isothermal halo fits to observed rotation curves. 
    The power spectra (Fig.~\ref{fig:collect_Spectra}) are evaluated inside the boxes delineated by the dashed lines. The large-scale gas flows, which convect the core-forming fluctuations in our model, are assumed to move inside these regions. 
    Of the 26 LITTLE THINGS galaxies for which isothermal mass models are available~\cite{2015AJ....149..180O}, we exclude DDO 46, DDO 110, and NGC 3738, which have unrealistically large inferred virial masses (${M}_{\text{vir}}{\simeq 10}^{{12}} {M}_\odot$ or larger). Additionally, we exclude Haro~36, which exhibits very large uncertainties on its isothermal core radius ($r_c =8.4\pm 12.17 \text{\ kpc}$).
     For each selected galaxy, the observational beam size is indicated by the dot in the lower-left corner of each panel.}
    \label{fig:maps}
\end{figure}
\begin{figure}
    \centering
    \includegraphics[width=\linewidth]{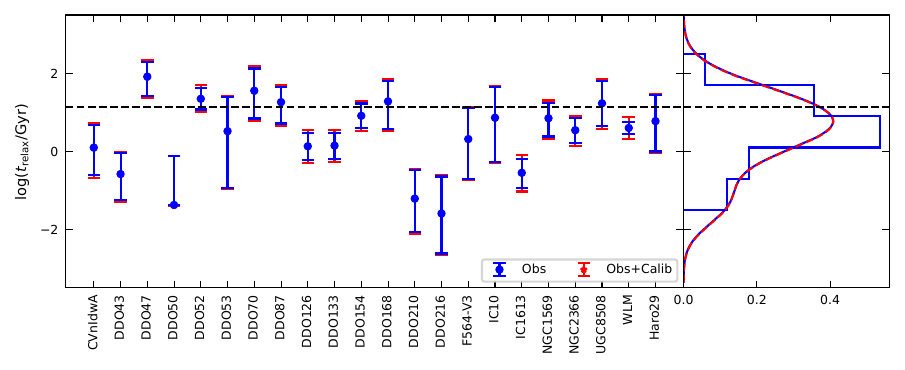}
    \caption{Distribution of relaxation times at the core $r_c$ of isothermal halo fits to observed rotation curves, relative to the Hubble time (horizontal line). The right hand panel shows a histogram of $\log(t_{\rm relax})$ overplotted with a Gaussian Kernel Density Estimation (KDE). It illustrates the clustering of the relaxation times around a value somewhat smaller than a Hubble time, which 
    may reflect a prolonged quasi-steady core formation process. The error bars  incorporate uncertainties on the index and normalization of power law fits to the 2D power spectra; on quantities derived from the mass model ($r_c$, $v_c$, $v_r$); on the effective density $\rho_0$; and on the minimal wavenumber $k_m$, corresponding to the largest fluctuation scales. Taken together, these are referred to as observational (Obs) errors, to distinguish them from the calibration errors (Calib) that result from using  the simulations and projection model to transform the 2D spectra into 3D ones (Section~\ref{sec:simps}).  A summary of the error calculation and propagation is provided in Section~\ref{sec:errors}, where Table~1 lists all  quantities entering 
    equation~(\ref{eq:relax}), with errors, for all galaxies.}
\label{fig:histo}
\end{figure}

Images showing the density contrasts 
$\Sigma ({\bf R})/\Sigma (R)$
in maps of the galaxy subsample
for which dynamical modelling is available, are shown in Fig.~\ref{fig:maps}, along with the 
relaxation times. As can be seen, the latter are 
of the order of a Hubble time or less for most galaxies. 
This is explicitly illustrated in Fig~\ref{fig:histo}, 
where the relaxation times are seen to cluster around a value
somewhat smaller than a Hubble time. The conspicuous exception 
is DDO~47, which has an anomalously large 
$t_{\rm relax}$. But as discussed in Section~\ref{sec:slopes}, 
it may also have a steep inner cusp.

\begin{figure}
    \centering
    \includegraphics[width= 13 cm]{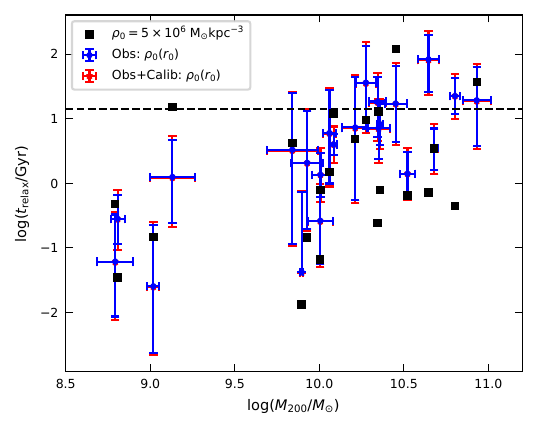}
    \caption{The relaxation time as a function of virial mass,  defined 
    as the mass within a radius enclosing 200 times the critical density. The points with error bars (labelled as in Fig.~\ref{fig:histo}) 
    refer to calculations using \(\rho_0 (r_0)\) as given by equation~(\ref{eq:rho0}), which reflects a decrease in atomic gas density in larger systems (cf. Fig.~\ref{fig:rho0}). The Spearman rank correlation coefficient is 
    $0.61 \pm 0.14$,  and the Pearson coefficient is $0.6 \pm 0.14$. For comparison, we also plot the relaxation times calculated while keeping a constant fiducial \(\rho_0 = 5 \times10^6~M_\odot/{\rm kpc}^3\), which may be apt if an increasing molecular gas contribution (possibly including a CO-dark component) to the total gas density compensates for the decrease of atomic gas density with mass. The virial masses and their errors are as 
    estimated in Section~\ref{sec:errors} (item 6).} 
\label{fig:relaxM200}
\end{figure}

We do not, of course, know the dynamical history of these galaxies; what we measure is the relaxation time  inferred from current parameters. 
Strictly speaking, it is related to future dynamical evolution. 
However, again, unless the gas fluctuations can be characterized 
by a perfectly stationary stochastic process, 
all crucial parameters in equation (\ref{eq:relax}) may 
collude towards smaller 
relaxation times, and thus more efficient cusp/core transformation  
in the past: the fluctuation levels (as measured by $P(k_m)$) should have been 
larger; the gas density $\rho_0$ should also have been larger; 
furthermore, the velocities should also have been 
smaller if the galaxy masses increased through past mergers. 
In this sense, the relaxation times calculated here are to be taken as upper limits. 
Alternatively, one can interpret the clustering of the results around a 
Hubble time as follows: suppose the process of stochastic fluctuations is indeed 
strictly stationary as the model assumes, 
then a core of the order of that observed is expected to form over a Hubble time, 
and consequently have an associated relaxation time of that order (as is the case with the simulation, where the relaxation time at the core radius at the end of the simulation was calculated to be of order a Hubble time~\cite{HEZFRC}).

However, we find a clear correlation between the relaxation time and the virial mass of the system, as shown in Fig.~\ref{fig:relaxM200}. Such scaling is expected, as the relaxation time increases as the cube of the velocity. This dependence arises from a $v_{p}^{2}$ term (due to larger particle kinetic energies in deeper potential wells) and a $v_{r}$ term (reflecting the fact that gravitational encounters between gas fluctuations and dark matter particles are weaker at larger relative velocities). Assuming a Tully-Fisher or Faber-Jackson type relation ($v \sim M^{1/4}$), the relaxation time scales with mass as $v^3 \sim M^{3/4}$.

For the points displayed with error bars in Fig.~\ref{fig:relaxM200}, the increase in $t_{\rm relax}$ with mass is further reinforced by the gas density scaling as $\rho_0 \sim 1/r_0$, where $\rho_0$ is given by equation~(\ref{eq:rho0}) and displayed in Fig.~\ref{fig:rho0}. This arises from the well-known mass-radius relation for HI gas and leads to a lower density in more massive systems. However, the HI density scaling may not universally hold; the total gas density may be nearly constant, or even increase with galaxy mass, when accounting for the increasing fractions of molecular gas in more massive galaxies. This may arise from an ascending molecular-to-atomic gas ratio~\cite{HIH22016A&A, HIH22018MNRAS}, potentially enhanced by undetected CO-dark molecular gas~\cite{COD2013ARA&A, DCO2017ApJ}. Thus, for more massive galaxies in particular, our fiducial estimates of the relaxation times (displayed in Figs.~\ref{fig:denscont} and~\ref{fig:histo}) should be taken as upper limits if a significant molecular component is present. For comparison, we also show the relaxation times evaluated while assuming a constant $\rho_0 = 5 \times 10^6 M_\odot/{\rm kpc}^3$ in Fig.~\ref{fig:relaxM200}, which are indeed generally shorter at higher masses.

On the other hand, the potential limitation of forming cores at higher mass may be compounded when accounting for the formation of dominant central stellar systems within haloes evolving in a cosmological 
setting~\cite{2014MNRAS.437..415D, 2016MNRAS.456.3542T, Maccio2020MNRASBH, CoreonFIRE2020MNRAS}. A related issue concerns whether cusps may reform during mergers, as cusps can represent equilibria of the merging process~\cite{1998MNRAS.293..337S, 2008ApJ...681.1058E}; massive haloes that grow through cusp-reforming processes may not maintain early-formed cores~\cite{2015MNRAS.454.2092O}. They are also less likely to form new ones given their longer relaxation times, unless they were subject to much higher density and potential fluctuations than those inferred here.

In summary, relaxation times of the order of a Hubble time or less indicate that past relaxation was sufficient to form significant cores in the present sample of galaxies. However, extrapolating the correlation in Fig.~\ref{fig:relaxM200} suggests that core formation through feedback-driven potential fluctuations may become less effective in more massive systems, particularly when assuming no significant contribution from molecular gas fluctuations. As a conservative conclusion, our results demonstrate that stochastic potential fluctuations can be an entirely viable mechanism for explaining halo cores within the standard CDM scenario, at least in systems with up to $M_{\rm vir} = 10^{11} M_\odot$ that have not undergone recent cusp reformation. Further analysis of imminent observations, with larger, deeper, and higher-resolution samples, may track the genealogy of core formation and possible cusp reformation, potentially elucidating the observed diversity in the central mass distribution of galaxies.



\section*{Methods}\label{sec11}
\setcounter{section}{1} 

\subsection{Sketch of the theoretical model 
leading to the relaxation time (equation~\ref{eq:relax})}
\label{sec:model}

In this section, we outline the relationship between density and potential fluctuations, as well as between spatial and temporal ones, within our framework. We then show how the fluctuations drive a random increase in the velocity dispersion of particles forming a halo cusp. In the process, we delineate the simplifying assumptions of the model. 
We finally discuss the physical picture envisioned and verify the association of the relaxation time with core formation.

Following~\cite{EZFC}, we first consider a time-snapshot  
of a gaseous medium with average density $\rho_0$. 
The gas is driven by an energy source,
causing large scale flows and 
an associated cascade of density and potential fluctuations 
within 
a volume $V$. To characterize this, we expand  
the density contrast $\delta = \frac{\rho({\bf r})}{\rho_0} - 1$ and 
potential $\Phi$ over a volume $V$, so that
\begin{equation}
\Phi  ({\bf r}) = \frac{V}{(2 \pi)^3}  \int \Phi_{\bf k} e^{-i {\bf k \cdot r}} d {\bf k},
\label{eq:phirk}
\end{equation} 
and 
\begin{equation}
\delta ({\bf r}) = \frac{V}{(2 \pi)^3} \int \delta_{\bf k} e^{-i {\bf k \cdot r}} d {\bf k}.
\label{eq:rhork}
\end{equation} 
The Fourier amplitudes are related through the Poisson equation, 
$\nabla^2 \Phi= 4 \pi G \rho_0 \delta$, by 
$\Phi_{\bf k}  = -4 \pi G \rho_0 \delta_{\bf k}  k^{-2}$, 
where $k = \vert {\bf k} \vert$.
The power spectrum  of density fluctuations is 
${P}({\boldsymbol k}) = V \langle |\delta _{\boldsymbol k}|^2 \rangle$, and that of the corresponding 
force fluctuations is 
${P}_F ({\bf  k}) = (4 \pi G \rho_0)^2 k^{-2} {P}(\bf k)$. 
The averages here are strictly speaking ensemble averages, but they may also 
be considered as average over all snapshots 
(in which case the process is said to be ergodic).  
The fluctuations $\delta ({\bf r})$ 
are statistically isotropic if 
$P({\bf k}) = P (k)$, 
implying that the  
variance in different spatial directions is the same: 
$\langle k_x^2 \rangle = \langle k_y^2 \rangle = \langle k_z^2 \rangle = \frac{1}{3} \langle k \rangle$.

If the fluctuations are described by time independent statistics
(a stationary stochastic process), 
the power spectra converge to the same values $P ({\bf k}) = V~\vert\delta_{\bf k}\vert^2$, 
for all snapshots, as $V$ increases. 
If the statistics of the fluctuations are independent 
of the point at which they are measured (a spatially stationary, or homogeneous, process), 
the spatial correlation functions are Fourier transforms of the power spectra.
In particular, the force correlation function is  
\begin{align}
\langle {\bf F} (0) \cdot {\bf F} (r)\rangle & = \frac{1}{(2 \pi)^3} \int  {P}_F ({ k})  \frac{\sin (k r)}{k r} 4 \pi k^2 d k.
\label{eq:FCFS}
\end{align}
To transform this into a time correlation function, we employ 
the 'sweeping' approximations long established in 
turbulence theory~\cite{Taylor1938,Kraichman1964,Tennekes1975}: if 
each mode with wavenumber $k$ is transported (advected) 
with velocity $v_k$, such that $r_k (t) = v_k t$, one can 
rewrite the above equation as 
\begin{align}
\langle {\bf F} (0) \cdot {\bf F} (t)\rangle & = \frac{1}{(2 \pi)^3} \int  {P}_F ({ k})  \left[\int p (v_k) \frac{\sin (k~v_k t)}{k~v_kt} d v_k \right] 4 \pi k^2 d k.
\label{eq:FCFT}
\end{align}
where $p (v_k)$ is the probability distribution function of $v_k$. 
The derivation of this equation 
and application to the case of Maxwellian $p$ was studied in~\cite{EZFCH}.
Here we follow~\cite{EZFC} and \cite{HEZFRC} in assuming
that the spectrum of fluctuations is swept 
by large scale fluid flows with a single characteristic 
speed $v_r$; so that $p (v_k) = \delta (v - v_r)$, 
and equation~(\ref{eq:FCFT}) reduces to~(\ref{eq:FCFS}) with $r(t) = v_r t$. 
In this way the spatial statistics of the fluctuating field are  
transformed into the time domain 
by being transported, frozen in, 
by large scale  flows with  characteristic speed $v_r$. 
The ergodic and sweeping assumptions were tested
against the results of the numerical simulation primarily used 
here in~\cite{HEZFRC} (Section 3.3). 

Using the Newtonian equations of motion, 
one may deduce that, after time $T$, the gas fluctuations  induce 
velocity changes on a dark matter particle 
with variance 
\begin{equation}
\langle (\Delta v_p)^2 \rangle = 2 \int_0^T (T - t)  \langle {\bf F}(0) \cdot {\bf F}(t) \rangle d t. 
\label{eq:corrsp}
\end{equation}
This is  proportional to the kinetic energy injection
from the fluctuating gas to the halo particles. 
Note that as $P_F$ scales as the density squared, so will the force 
correlation function from equation~(\ref{eq:FCFS}), 
and consequently so will the kinetic energy injection rate.   

For a power law spectrum $P_F \sim k^{-(n+2)}$ (with 
$n$ the index of density spectrum), one may evaluate this energy, 
assuming maximal and minimal   
cutoffs, $k_x \gg k_m$. When the diffusion limit 
($v_r  t \gg k_m^{-1}$) is taken, equation~(\ref{eq:relax}) 
follows by setting
$\langle (\Delta v_p)^2 \rangle = v_p^2$ 
as is usual in calculations of the 
relaxation time. 

As the correlation function in (\ref{eq:corrsp})  should be understood as taken along the path of a dark matter particle, 
$v_r$ is the typical 
speed of the large scale gas flow with respect to the particle. 
We follow~\cite{EZFC} in associating  this 
with the larger characteristic speed of repeated outflows and 
inflows confined within  spatial scale   
$d \simeq V^{1/3}$.  

The  physical picture envisioned is that of 
persistent fountain flows
convecting  a cascade of multiscale fluctuations 
within $V$, rather than outflows permanently 
exiting $V$. 
In this context, $v_r$
refers to an average speed corresponding to the flows being in 
energetic equilibrium with the gravitational potential within $d$. 

\begin{figure}[H]
    \centering
    \includegraphics[width= 12 cm]{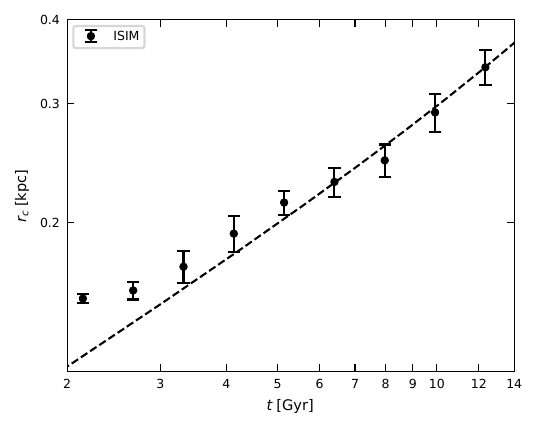}
    \caption{Comparison of the time evolution of the core radius in ISIM, as obtained from an isothermal fit with core $r_c$ (with fitting error) of the halo density profile, with the timescale obtained from equation~(\ref{eq:relax}). Following~\cite{HEZFRC}, we use
    $\rho_0 = 10^6 M_{\odot}~\rm kpc^{-3}$, and spectral parameters $P (k_m) = 4~\rm kpc^3$, and $n= 2.3$, also in line with values found in~\cite{HEZFRC} (Fig.~ 4; and already predicted in~\cite{EZFC}). We set  $v_p = v_c$, as assumed in the calculations evaluating the relaxation time for the galaxy sample (thus, $v_c$ is set to the rotation speed at $r_c (t)$).  Following refs.~\cite{EZFC, HEZFRC}, the velocity of large scale gaseous flows relative to the halo particles is 
    estimated  in terms of the average dark matter density within scale $d/2$ as 
    $v_r = d~ \sqrt{G \bar{\rho}_{\rm DM}(d/2)}/2$. 
    Here $d = 5~{\rm kpc}$ is the diameter inside which the large scale flows carrying the fluctuations are assumed to be confined, and outside which little outflow or inflows occur (its approximate value can be estimated from Fig.~2 of ref.~\cite{HEZFRC}). This is the region where the fluctuations leading to core formation are assumed to occur, and thus also defines the size of the box over which the power spectrum is taken.}
\label{fig:core_relax}
\end{figure}

As the  volume $V$ inside which  this takes place is finite, an ideal statistical 
stationarity process as defined above cannot strictly apply, even in principle (due to random variations 
between snapshots). 
But a weaker condition of quasi-stationarity may still apply if the 
power spectra display a well defined mean and variance over time. 
This is reflected in 
Fig.~\ref{fig:collect_Spectra}  by the limited 
variance in the simulated spectra.
Physically,  'bursty' feedback  can prompt a quasi-stationary
process if it leads to similar (in terms of energy input and effect on the gas) 
consecutive episodes 
of fountain flows within $V$, 
driving density and mass fluctuations that display similar statistics over time.
In an isolated simulation this leads to the stability of the power spectra over time
(as in Fig~1 and also Figs. 4 and 5 of ref.~\cite{HEZFRC}).  
In cosmological simulations, the situation is complicated by 
mass changes arising from accretion and mergers. In that case, the process 
may be considered piecewise quasi-stationary over limited time (as discussed in 
Section~\ref{sec:cosmo} below).

In ref.~\cite{HEZFRC}, a 
detailed model for the evolution 
of the central dark matter distribution under the influence of the gas  
potential fluctuation was presented.  It fit 
the numerical results with no free parameters 
unconstrained by the simulation. 
As those parameters are not available from observations, 
we here associate  the formation time of a core with radius 
$r_c$ with the relaxation time at $r_c$. This is 
estimated by setting the dark matter velocity to the circular 
velocity at $r_c$ in equation~(\ref{eq:relax}).  
As mentioned in the main text, the procedure may be justified by the fact $t_{\rm relax}$ is generally 
an increasing function of radius. 
We tested this criterion against the simulation 
for the same fiducial parameters used 
in ref.~\cite{HEZFRC}. 
The result is shown in Fig.~\ref{fig:core_relax}.

\subsection{The surface densities of observed galaxies}
\label{sec:sudens}

To produce the HI surface density maps of the LITTLE THINGS galaxies, we used the HI flux images created using the procedures described in Section~3 of ref.~\cite{2012AJ....144..134H}.~\footnote{The data cubes and constructed images are taken from the NRAO repository for the LITTLE THINGS survey: \href{https://science.nrao.edu/science/surveys/littlethings}{https://science.nrao.edu/science/surveys/littlethings}.}
 The HI data cubes, in the image-domain, are  thus created with the ROBUST = 0.5 weighting scheme in the AIPS task IMAGER. The cubes are then cleaned with a threshold flux level of two times the RMS noise, as determined in the line-free channels. The data cubes are furthermore  corrected for primary beam attenuation, and then de-projected using 
 the photometry parameters (galactic centre, position angle, and axial ratio) 
determined from the optical V-band morphology \citep[table 6]{2006ApJS..162...49H, 2012AJ....144..134H}. We compared the results with 
the original maps, with no de-projection. As  the inclinations of the LITTLE THINGS galaxies are generally less than 30 degrees the differences in final surface density profiles were not generally large,  although in some 
cases the central densities are larger 
by up to a factor of two in the original maps due to larger gas columns resulting from inclination effects.  
The use of the de-projected maps matches the requirements for evaluating the 3D density (Section~\ref{sec:rho0}) and projecting power spectra (Section~\ref{sec:project}), which assume cylindrical symmetry along a projection axis normal to the image.

The HI mass is obtained from the  integrated HI flux ($S$) in the robust weighted maps using
\begin{equation}
 M_{\rm HI} = 235.6 \times
 \left(\frac{D}{\rm Mpc}\right)^2 \sum_{ch} \left(\frac{S_{ch}}{\rm mJy}\right) \left(\frac{\Delta V}{\rm km~s^{-1}}\right) M_\odot,  
\end{equation}
where $D$ is the distance to the galaxy and $\Delta V$ is the velocity channel (ch) width.  The surface density is then given by:
\begin{equation}
\Sigma_{\rm HI} = M_{\rm HI}/A_{pixel},     
\end{equation}
where $A_{\rm pixel}$ is the pixel area in ${\rm kpc^2}$. The HI surface density  profiles $\Sigma(R)$ are then measured in annuli of width equal to the HI beam size.

\begin{figure}
    \centering
    \includegraphics[width= \linewidth]{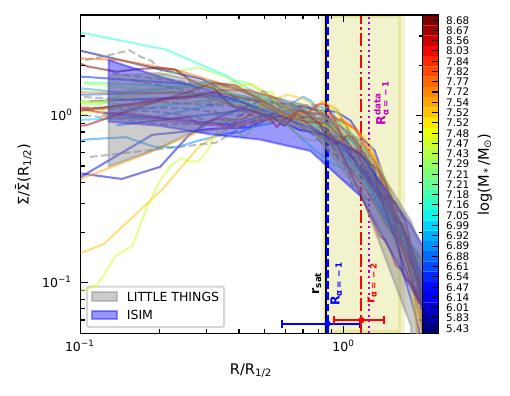}
    \caption{Collected gas  
    surface density profiles in the galaxy sample and  reference simulation (ISIM). The horizontal axis is scaled to the half mass radius $R_{1/2}$ and the vertical axis to the average density within it. For comparison with data, where observational sensitivity imposes a maximal cutoff radius, the total gas mass in the simulation 
    is assumed to be enclosed within radii where  $\Sigma > 10^{19}$ ${\rm atoms/ cm^{2}}$ (with $R_{1/2}$ enclosing half this mass).   
    The shaded areas delineate  the dispersion in the 
    galaxy curves and  simulation snapshots. 
    The vertical solid line corresponds to the 
    energy injection saturation radius in ISIM ($r_{\rm sat}$), beyond which a steep decrease in density prevents energy transfer from the gas fluctuations to the particles forming  the halo cusp; the dashed line denotes the radius beyond which the surface density decreases with logarithmic slope steeper than $\alpha = -1$ ($R_{\alpha = -1}$), with horizontal bars denoting the dispersion 
    (arising from variations between snapshots and slope range considered, as described in the text); the  dashed-dotted line  denotes  a corresponding criterion for the 3D density, with $\alpha \le -2$. Within the estimated uncertainties, the measures involving the slopes are consistent, and close to both $r_{\rm sat}$  and $R_{1/2}$ (corresponding to unity on  the x-axis). 
    They are also consistent with 
    $R^{\rm data}_{\alpha = -1}$ of the galaxies (with mean denoted by the dotted vertical line and dispersion with the yellow shaded region).}  
    \label{fig:SurfaceDensl}
\label{fig:collect_project_dens}
\end{figure}

 The collected results for all galaxies in the sample are shown 
in Fig.~\ref{fig:SurfaceDensl}, along with the surface density obtained from the simulation ISIM, which is used to calibrate the fluctuation power spectra as described below.  In presenting the results we scale 
our curves to the half mass and density, as this exploits the near homology reflected in the HI mass-radius relation~\cite{2016MNRAS.460.2143W}. This facilitates 
the comparison of the profiles of systems with different 
gas masses.

\subsection{The characteristic density $\rho_0$}
\label{sec:rho0}

Equation~(\ref{eq:relax}) 
involves the 3D average gas density $\rho_0$. 
If the gas is homogeneous on large scales, $\rho_0$ 
tends to a definite value when measured over increasing scales $r_0$.  
In a realistic gas distribution, on the other hand, 
the density generally decreases 
with radius,  falling off steeply beyond a certain scale. 
In this case,  the relevant effective radius $r_0$ 
is one inside 
which $\rho_0 (r_0)$ is large enough to ensure 
significant energy transfer from the fluctuations in the gas to the 
particles making the dark matter cusp. 
This radius was estimated for the simulation 
studied in ref.~\cite{HEZFRC} by measuring the actual energy 
transfer from the gas to the dark matter within 
progressively larger radii (cf. Fig.~7 of that work). 
In this case, it was found that it saturated beyond radius $r_{\rm sat}= 1.3~{\rm kpc}$, which 
may in this case be identified
with $r_0$.
The scale $r_{\rm sat}$ 
cannot be deduced this way from 
data, but it may nevertheless be connected to 
the properties of the observed density 
profile. 

In ref.~\cite{HEZFRC} it was noted that 
$r_{\rm sat}$ 
approximately matched the radius at which the  
density fell off steeper than $\rho \sim 1/r^2$. 
Since, as discussed in Section~\ref{sec:model}, the rate of energy transfer  
from the gas fluctuations to the dark matter scales as the square of the density, 
this corresponds  to a transfer rate 
rapidly declining at least as $1/r^4$. 
Furthermore,  an $\sim 1/r^2$ profile  corresponds to surface density 
$\Sigma \sim 1/R$, when projected, we may expect the  projected
radius where the surface  density starts falling off as 
$1/R$ to also approximate $r_{\rm sat}$. 
We checked this by examining the 2D and 3D density slopes in our fiducial simulation. For each snapshot, we calculated the logarithmic slope 
of the 3D gas density $\alpha = \frac{d \ln \rho}{d \ln r}$ and  evaluated
the radius at which $\alpha \le b$. To reduce 
noise, we repeat the calculation for a range of values within
$-2.5 \le b \le -2$.
We then average the resulting cutoff radii over different 
$b$ and over simulation snapshots,
to obtain the cutoff radius 
$r_{\alpha = -2}$. 
We do the same for the surface density, except that 
in this case $-1.5 \le b \le -1$, to obtain $R_{\alpha = -1}$. 
The results are shown in Fig.~\ref{fig:SurfaceDensl} by the 
vertical lines. As can be seen, the two measures 
are consistent, within the error bars, and close to $r_{\rm sat}$ (indeed $R_{\alpha  = -1}$ turns out to better 
approximate $r_{\rm sat}$).

Fig.~\ref{fig:SurfaceDensl}, also shows that all 
three aforementioned 
measures match the projected half mass radius 
(at unity). They are likewise close to 
$R_{\alpha  = -1}$ for the galaxy 
gas density profiles
(denoted by $R^{\rm data}_{\alpha  = -1}$). 
The latter is  shown by the vertical dotted line,  with dispersion due to scatter between galaxies delineated 
by the surrounding vertical shaded region. 
Because the half-mass radius \(R_{1/2}\) falls within this scatter, we will use it to define $r_0$ and $\rho_0 (r_0)$.  Choosing \(R_{1/2}\) over \(R_{\alpha =-1}\) is advantageous because it is readily defined and evaluated,  avoiding the profile density 
differentiation that introduces significant uncertainties.

\begin{figure}
    \centering
    \includegraphics[width= 12 cm]{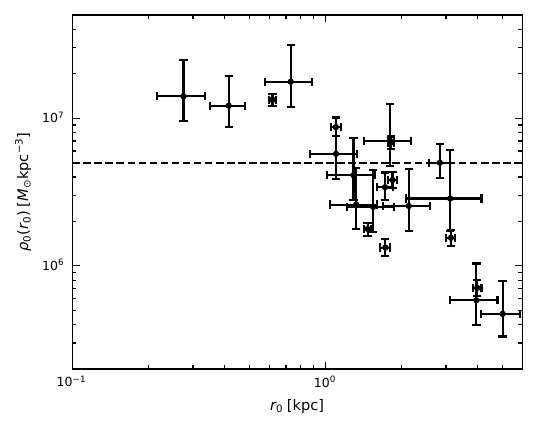}
    \caption{The characteristic gas densities $\rho_0 (r_0)$ entering into equation (\ref{eq:relax}) as calculated from equation (\ref{eq:rho0}). The trend of decreasing density with radius is consistent with the mass radius relation of HI gas~\cite{2016MNRAS.460.2143W}. The Spearman rank correlation coefficient is $-0.81 \pm 0.07$ and the Pearson coefficient is $-0.84 \pm 0.07$. The dashed horizontal line
    denotes the reference constant density 
    used in Fig.~\ref{fig:relaxM200}, to take into account a possible molecular gas contribution.}
\label{fig:rho0}
\end{figure}

In estimating the 3D density at $R_{1/2}$, it is simplest
to assume cylindrical symmetry, as the observed 
surface density is defined in terms of projections
on cylindrical shells with radii $R$ 
on the image planes, along the normal axis $z$ 
(an analogous argument, assuming the density is stratified along spheroids 
leads to similar results). 
The density profile of the gas is thus 
assumed to be of the form $\rho (R, z) = \rho (R) \rho\ (z)$,
which is constant along
cylindrical shells with axis ratio $z/R = q \le 1$. 
For the flattening of dwarf irregular galaxies 
a  characteristic value is  
$q \simeq  0.4$~\cite{1988PASP..100..344V, 2013MNRAS.436L.104R, 2017MNRAS.465L..49J}. 

The enclosed mass is given in terms of 
the average density inside the shells  by 
$M (R, z) = M (R, q R) = 2 \pi q R^3 \bar{\rho} (R, q R)$, 
The average density at the half projected mass radius $R_{1/2}$ is 
\begin{equation}
  \bar{\rho}_{1/2} (\eta, q) =  \frac{\eta~M_{1/2}  (R_{1/2}, q R_M)}{2 \pi q R^3_{1/2}}.
\end{equation}
Here $M_{1/2}$ is the measured projected 
mass at $R_{1/2}$, and the factor 
 $\eta = M (R_{1/2}, q R_{1/2}) / M (R_{1/2}, q R_M)$ accounts for the fact that $M_{1/2}$ is defined in terms 
of the whole projected gas column density up to a total (maximum) height $z = q R_M$
normal to the image planes, 
while the average 3D density is defined for mass inside  $z = q R_{1/2}$.
For homogeneous density distributions (with sudden cut-off at ($R_M, q R_M$)), 
$\eta = R_{1/2}/ R_M = 1/\sqrt{2}$, and is generally closer to unity when the 
density  gradually decreases outwards from the centre.~\footnote{For example, if $\rho \propto e^{- \vert z \vert /q}~ e^{-R}$, as $R_M \rightarrow \infty$ 
the mass ratio 
is $\frac{M (R,~q R)}{M_\infty} = (1- e^{- \vert z \vert /q}) 
[1 - e^{-R} (1+R)]$; the projected half mass radius 
is at $x_{1/2} =1.68$ and 
$\eta = 0.81$. If the distribution in the $z$ direction 
is replaced by a Gaussian, the corresponding ratio is 
$\eta = 0.9$.}   
We then define $\rho_0 (r_0) = \bar{\rho}_{1/2} (\eta, q)$ 
and an effective spherical radius 
\begin{equation}
r_0 = \left(\frac{3 q}{2 \eta}\right)^{1/3} R_{1/2},
\label{eq:r_o}
\end{equation}
such that 
\begin{equation}
   \rho_0 (r_0) = \frac{M_{1/2}}{\frac{4}{3} \pi r_0^3}.
   \label{eq:rho0}
\end{equation}
Note that $r_0 \simeq R_{1/2}$; 
 e.g., $r_0 = (0.8)^{1/3} R_{1/2}$ for $q=0.4$ and 
$\eta =0.75$. 

The  densities $\rho_0 (r_0)$, thus calculated, 
are shown in Fig.~\ref{fig:rho0}.  In this plot, as in all 
calculations using equation (\ref{eq:rho0}), the mean value of $q$ and 
the uncertainty on it is taken, for each galaxy,  
from ~\citep{2017MNRAS.465L..49J}, and a flat prior on 
$\eta$, such that $0.71 \le \eta \le 0.95$.  In calculating the mass, 
the mass of HI is multiplied 
by a factor 1.4 to account for helium and other elements.  

\subsection{Simulated power spectra and calibration}
\label{sec:simps}

\subsubsection{Isolated simulation (ISIM~\cite{HEZFRC})}
\label{sec:ISIMcalib}

\begin{figure}[H]
    \centering   
    \includegraphics[width= \textwidth]{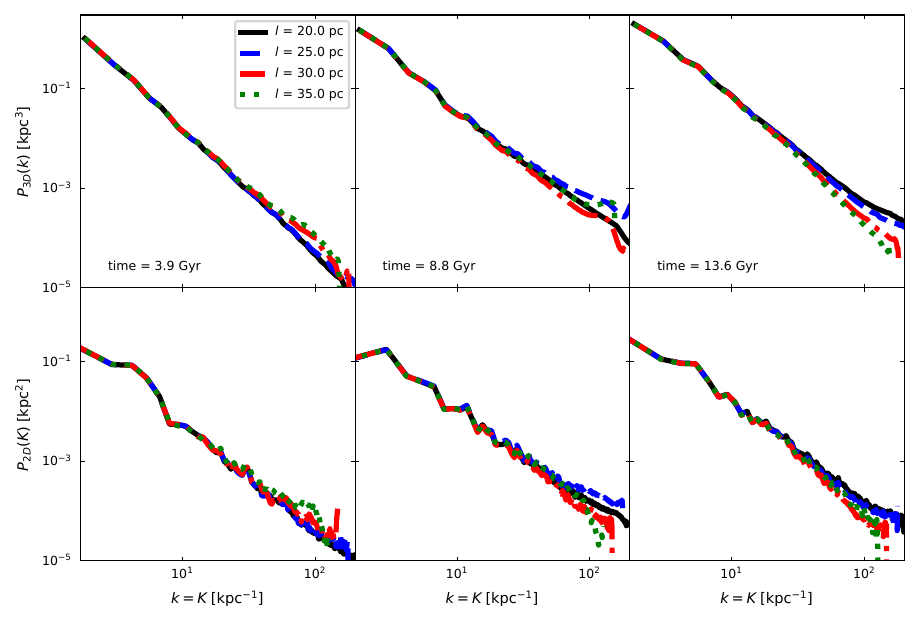}
    \caption{Power spectra in our reference simulation (ISIM)  
    for different FFT pixel sizes, with 
    3D spectra on the upper panels and 2D ones on the lower panels.}  
\label{fig:pixels}
\end{figure}

\begin{figure}
    \centering   
    \includegraphics[width= \textwidth]{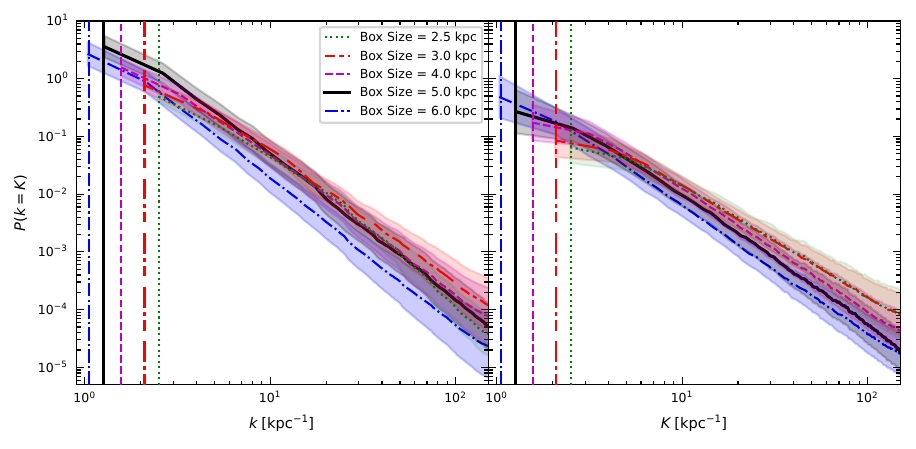}
   \caption{Power spectra (3D on the left and 2D on the right) over different FFT box sizes in ISIM. For each box, averages $\langle\vert\delta_k\vert^2 \rangle$ are taken over 200 snapshots between 4~Gyr and a Hubble time, and the dispersion is shown by the shaded regions. Vertical lines denote the small-wavenumber limit imposed by the box size (with larger spatial fluctuation scales captured by larger boxes). In choosing the box size, a compromise is made between capturing the largest fluctuation scales and avoiding regions where feedback-driven fluctuations are negligible. The outer limit of the region relevant to our model can be estimated from the temporal stability of the enclosed gas mass shown in Fig.~2 of~\cite{HEZFRC}. This suggests that the box size should not greatly exceed $d/2 \approx 2$~kpc. The smallest box size corresponds to slightly less than twice 
   the effective density scale radius $r_{\rm sat} = 1.3~\rm kpc$.}
\label{fig:psbox}

\end{figure}

We are interested in the power spectra of 
fluctuations in the density contrasts, 
as their normalization and tilt  enter into equation~(\ref{eq:relax}).
We define these in 2D and 3D through 
 $\delta_{2D} = \Sigma ({\bf R})/\Sigma(R) -1$ 
and $\delta_{3D} = \rho ({\bf r})/ \rho (r)  -1$. Here 
$\Sigma (R)$ and $\rho(r)$ refer, respectively, to averages over circular rings 
and spherical shells at cylindrical radii $R = \vert {\bf R} \vert$, and 
spherical radii $r = \vert {\bf r}  \vert$. 
 Over a surface $S = d^2$ and volume $V = d^3$ 
the corresponding Fourier modes are 
\begin{eqnarray}
\delta_{\bf K}  = \frac{1}{S} \int \delta ({\bf R})~e^{i {\bf K} \cdot {\bf R}}  d {\bf R},\\
\delta_{\bf k} = \frac{1}{V} \int \delta ({\bf r})~e^{i {\bf k} \cdot {\bf r}}  d {\bf r}. 
\end{eqnarray}
In practice, the power spectra are calculated through discretization 
of these equations and applying a fast Fourier transform (FFT) 
at a resolution defined by a pixel  size $l \ll d$.  
In the convention used here, the power spectra are computed, at each snapshot, 
from  $P ({\bf K}) = S \vert \delta_{\bf K} \vert^2$
and $P ({\bf k}) = V \vert \delta_{\bf k} \vert^2$.
The isotropic spectra are obtained by averaging over shells
or rings around radial wavenumbers  $k$ and $K$ respectively. 

On the spectra inferred from  the simulation ISIM, we performed various 
convergence and robustness tests. 
In Fig.~\ref{fig:pixels}, we show the effect of pixel size $l$ over which the 
discrete Fourier transform is evaluated. Except for 
an expected resolution effect at large wave numbers, the results 
are stable. This robustness of the spectra 
at small wavenumbers will  be  confirmed in further tests discussed 
below in connection to the observations. As our  principal results depend on the normalization of the spectra at small 
wavenumbers---$P (k_m)$
in~(\ref{eq:relax})--- 
this property renders them also robust.  

We also tested the effect of inclination
on the 2D and 3D power spectra, by rotating the system relative 
to axis of angular momentum, 
and found the results to be insensitive to it.  This  
reinforces the assumption of isotropic fluctuations 
(cf. Section~\ref{sec:model}). 

In the above tests, we used the fiducial box size $d = 5~{\rm kpc}$ (so that $d/2 \simeq 2 r_{\rm sat}$), following ref.~\cite{HEZFRC}. Variation of the box size $d$ does not introduce very large differences as long as $d/2$ is large enough to incorporate the largest fluctuation scales, but not so large as to contain regions free of fluctuations---that is, beyond the region where episodic feedback-driven gas outflows and inflows are confined. From Fig.~2 of~\cite{HEZFRC}, this region satisfies $d/2 \gtrsim 2~{\rm kpc}$.

Thus, while in an idealized situation where fluctuations exist over an arbitrary volume $V$, the power spectra should converge as $V$ increases, when the fluctuations are confined to a finite $V$---as in a realistic simulation, and as assumed in the physical picture envisioned in our model (Section~\ref{sec:model})---the spectra will weaken as regions essentially free of fluctuations are incorporated. A compromise is therefore necessary between selecting box sizes that capture large fluctuation scales and avoiding the inclusion of significant fluctuation-free regions.

With this in mind, in Fig.~\ref{fig:psbox} we 
show results for a range of $d$ that generously encompasses all relevant 
scales, from $d = 2.5~\text{kpc}$ (which is smaller than the effective density 
scale) to $d = 6~\text{kpc}$ (which encompasses a volume larger than that 
within which fluctuations are expected to be significant). The spectra in 
this latter case indeed fall below the other lines, which are 
more consistent within the ranges allowed by the box sizes and the 
uncertainties shown by the shaded regions. Nevertheless,  as we will now see, for the 
purpose of transformation from 2D to 3D spectra, the differences are less 
pronounced; the ratios of the 3D to 2D spectra are on average less sensitive to 
the larger box size than the absolute values, and so the $6~\text{kpc}$ box 
will also be used for that purpose.

\begin{figure}
    \centering
    \includegraphics[width=\textwidth]{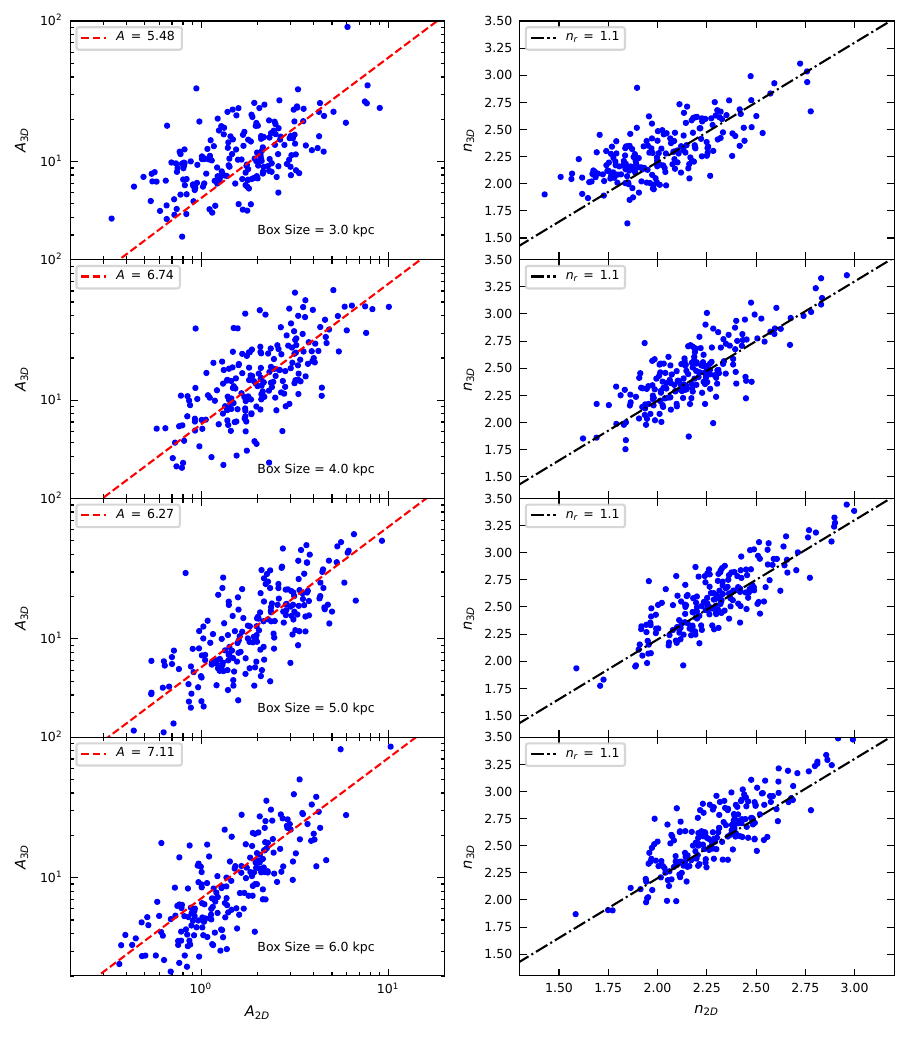}
    \caption{Calibration of the 3D power spectrum to the 2D 
    spectrum from the reference simulation ISIM. 
    We fit the 2D and 3D spectra in a scaling region 
    (as defined in text) with power laws (equations~\ref{eq:PPS}); the  
      panels show the correlation between the estimated 2D and 3D indices and normalizations  for different simulation snapshots. The numbers labelled $A$ refer to the ratios of the 3D to 2D spectra 
      at $k = 1~{\rm kpc}^{-1}$. The ratios of the power law indices are denoted by $n_r$.  
    Moving vertically, the panels correspond to different box sizes $d$ over which the spectra are evaluated (as labelled).}
    \label{fig:PScalib}
\end{figure} 

As $P (k_m)$ and $n$ in equation~(\ref{eq:relax})
refer to the 3D power spectrum, while 
only 2D spectra are evaluated from observations,  
we use snapshots from ISIM to calibrate the 3D spectra to the 2D ones. 
 In the following two subsections we 
extend the results to systems with different 
effective gas radial scale ($\simeq r_0$), by examining higher mass 
haloes from the FIRE-2 simulation and through a simple projection model. 

To extract spectral parameters, we define a power-law scaling domain:
starting at the midpoint between wavenumbers 
of 1 and 50 $\rm kpc^{-1}$, we calculate the logarithmic derivative, moving 
away from the midpoint until the slope deviates by 0.5 from the central value, defining this as the power law scaling 
region of the spectrum
(the relatively large deviation allowed takes into account fluctuations in the logarithmic derivatives due to noise).  
  Within this scaling window, we fit power laws of the form: 
\begin{eqnarray}
\nonumber 
P_{\rm 2D} = A_{\rm 2D} \left(\frac{K}{\rm kpc^{-1}}\right)^{-n_{\rm 2D}} {\rm kpc ^2} \\ 
P_{\rm 3D} = A_{\rm 3D}.\left(\frac{k}{\rm kpc^{-1}}\right)^{-n_{\rm 3D}} {\rm kpc^3}, 
\label{eq:PPS}
\end{eqnarray} 
where $A_{\rm 2D}$ and $A_{\rm 3D}$ are dimensionless
parameters. 
The results for box sizes between 3 and 6~kpc are shown in Fig.~\ref{fig:PScalib}. 
We find the 2D and 3D slopes to be 
nearly identical, with  slightly steeper 3D slopes. 
The amplitudes $A_{\rm 3D}$ are significantly larger 
than $A_{\rm 2D}$, but  their  ratio 
is consistent with 
being approximately constant.   
Taken together, these properties imply a quasilinear relation between $P_{\rm 3D}$ and $P_{\rm 2D}$, 
which 
may be interpreted in terms of the projection model 
described below (Section~\ref{sec:project}). 

The mean values of the indices are generally 
related by  
$n_{3D}  \equiv n \simeq 1.1 n_{2D}$.  
For  $d = 5~{\rm kpc}$, the fiducial value 
used in~\cite{HEZFRC}  and Fig.~\ref{fig:collect_Spectra},
the amplitudes are related  
through $A =A_{\rm 3 D}/A_{\rm 2D} = 6.27$. 
For the purpose of the transformation 
of the observed 2D spectra into 3D ones, we 
use the ratios of the amplitudes and slopes of all 
simulation snapshots and box sizes used, and estimate 
the mean values and the associated errors 
(by evaluating the dispersion). 
This gives
\begin{eqnarray}
\nonumber 
n_r = \frac{n_{3D}}{n_{2D}} = 1.13 \pm 0.08\\
A = \frac{A_{3D}}{A_{2D}} = 6.3 \pm 4.
\label{eq:calib}
\end{eqnarray}

\subsubsection{Comparison with cosmological simulations}
\label{sec:cosmo}

A  detailed examination of the mechanism studied here 
in the context of cosmological simulations is better suited for a separate study. Nevertheless, 
to investigate the presence, or lack thereof, 
of the essential features required by our model,
and to verify the calibration beyond the idealized simulation considered above, 
we examined the gas properties in haloes
identified in two different cosmological simulations for which data 
are publicly available, and which employed substantially 
different star formation and feedback recipes. 

This included several small 
haloes from the public release of the Auriga gravo-MHD 
cosmological zoom-in simulation~\cite{AURIGA2024}. 
The subgrid physics in this suite follows the prescription
of Springel \& Hernquist~\cite{SpringelHernq2003}. 
The gas is considered as a two-phase medium, with star formation occurring in the cooler component. 
The threshold for stochastic star formation is small 
(at number densities $0.13 {\rm cm}^{-3}$), 
and the process 
is regulated by a long timescale (2.2 Gyr), a stiff equation of state, and the removal of gas through large scale winds 
in proportion to the star formation rate. 
This makes for more quiescent star formation, with bursty episodes accompanied by large density 
fluctuations being less likely.

The 2D power spectrum of gas embedded in an example halo
from the Auriga Level~3 suite (resolution $\sim 30 {\rm pc}$) 
was already presented in Fig.~\ref{fig:collect_Spectra}. 
Relative to the observations and our fiducial simulation 
ISIM, the slope and normalization of the spectrum 
represent quantitatively and qualitatively  different gas properties; in terms of 
the level of the fluctuations and their distribution among scales, respectively.  
The central densities of haloes
from this  simulation were previously analysed, 
and density cusps, rather than cores,  were  invariably found~\cite{2019MNRAS.486.4790B,CuspAURIGA2023}. 
As we will see below, this is also the prediction of our model. 
Indeed, as we will find, the 3D gas spectra in Auriga lack extended power-law scaling regions. 
This reflects the absence of a scale-free cascade of density fluctuations that may be carried by larger scale flows, which is a central feature of our model (Section~\ref{sec:model}), In light of this, and the differences with the 2D spectra of the galaxies, we will not use the Auriga results in  verifying the projection model that calibrates 2D to 3D spectra (Section~\ref{sec:project}).

We also examined the properties of the 
gas in haloes in publicly released data from the 
FIRE-2 zoom-in simulations~\cite{FIRE-2-Pub2023ApJS}. 
These simulations have higher (pc scale) resolution 
and different recipe for the gas physics. 
In particular, the star formation threshold is far higher 
than in Auriga, and unlike the former is unregulated; 
its onset is set by the local Jeans scale and  
it occurs on the local cloud collapse time. This may lead to 
bursty episodic feedback, blowing holes and driving large scale 
flows, which  can carry large density fluctuation cascades 
of the type envisioned in our model. As already seen in Fig.~\ref{fig:collect_Spectra},
the spectra of the 2D fluctuations are also closer to  
those of the galaxies. We will thus use the results from the FIRE suite 
for verifying the spectral calibration process in Section~\ref{sec:project}. 

The issue of core formation in the FIRE simulations was first studied in~\cite{2015MNRAS.454.2092O}.  
Cores were found to be present in some haloes, though cusps were also found to reform 
through mergers,  an effect we do not consider here.  With the goal of  
verifying the projection and calibration model in mind, we  consider instead the gas properties
inside haloes where cores were identified in the 
FIRE-2 simulation~\cite{CoreonFIRE2020MNRAS}, and do this within time intervals 
that are large enough to evaluate a statistical 
sample of spectra but which are still small compared to the  timescale of halo mass evolution, which necessarily 
occurs through accretion and mergers in a cosmological context.~\footnote{ For completeness, we also examined gas fluctuations (or the lack thereof) in the publicly available ultra-faint dwarf galaxy halo (labelled $\rm m10v_{30}$ in ref.~\citep{2019MNRAS.490.4447W}), which has a small stellar-to-halo mass ratio of $3.7\times 10^{-5}$ at $z = 0$. (For comparison, the corresponding ratios for the cored FIRE-2 systems considered are $1.4\times 10^{-2}$, $10^{-2}$, and $3.9\times 10^{-3}$ for $\rm m11d$, $\rm m11e$, and $\rm m11q$, respectively.) The power spectra were found to be 
similar 
to those in Auriga, exhibiting a limited single power law scaling region and small 
normalization. This, coupled with a low gas density, leads to $t_{\rm relax} \approx 10^5$ 
Hubble times upon direct application of equation~(\ref{eq:relax}).}

This ensures quasi-stationarity during that period; 
by which we mean that the power spectra form a homogeneous sample,
such that the average remain stable, during the period considered, with modest scatter (Section~\ref{sec:model}). After some trials, we choose time intervals of 
$\sim 0.5~{\rm Gyr}$, during periods when no major mergers occur. 
Such a  time interval is significantly larger than  the timescale of the largest scale fluctuations ($t_f \sim (k_m v_r)^{-1}$).
The separation of timescales 
allows for multiple realizations of the 
fluctuation field while 
the global parameters of the system (like total mass) 
do not change significantly. Within the considered timescale, a  halo particle thus  
experiences fluctuations with the same quasi-stationary statistics. 
(This has analogy for example in the application of two body relaxation 
time on evolving haloes considered in~\cite{ZantRelaxCosm2006MNRAS}). 
The total effect of the gas fluctuations, as modelled here, 
may be considered as 
a cumulative effect of such quasi-stationary time intervals. 

\begin{figure} [H]
    \centering
    \includegraphics[width= 13. cm]{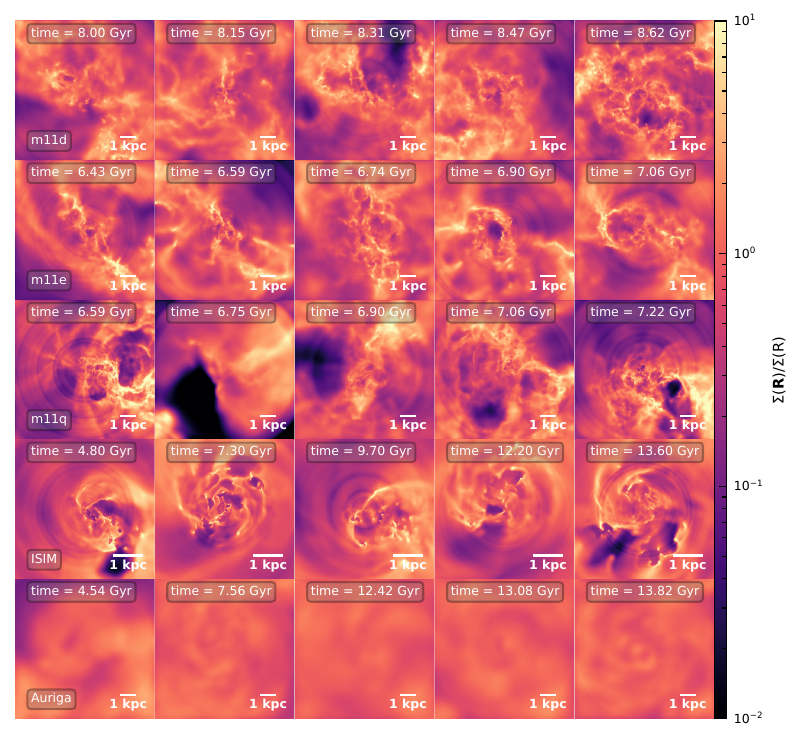}
    \caption{Projected density contrasts of the gas embedded in three haloes identified  in the FIRE-2 simulation Core Suite as hosting bright dwarf galaxies (top three panels, with identifiers as labelled in Table~1 of~\cite{FIRE-2-Pub2023ApJS}); the fiducial  isolated simulation (ISIM, fourth panel); and the halo from the Auriga simulation with spectrum shown in  Fig.~\ref{fig:collect_Spectra} in the bottom panel (halo ID 1 in Table~A1 of~\cite{AURIGA2024}, referred to as halo~1 in the text). The density fluctuations are qualitatively similar in FIRE and ISIM.  
    In the Auriga simulation, in contrast, only weak large scale asymmetries and spiral patterns are apparent, with no evident scale-free cascade, associated in Fourier space with an  extended power law spectrum. This is 
    reflected in the spectra of Fig.~\ref{fig:cospower}, where an additional Auriga halo is also considered.
    As the statistical properties encoded in the power spectra are similar in FIRE and the galaxies, we add the FIRE results to those of ISIM in verifying the power spectrum projection model (section~\ref{sec:project}).
    The time intervals for the FIRE simulations are accordingly chosen to be smaller to limit the time dependence arising from cosmological evolution and to enable comparison of the resulting relations between 2D and 3D spectra (Fig.~\ref{fig:FIREA2DA3D}) with those in the isolated ISIM (Fig.~\ref{fig:PScalib}).
} 
\label{fig:denscont}
\end{figure}

\begin{figure}[H]
    \noindent\centering
    \includegraphics[width= 0.49\textwidth]{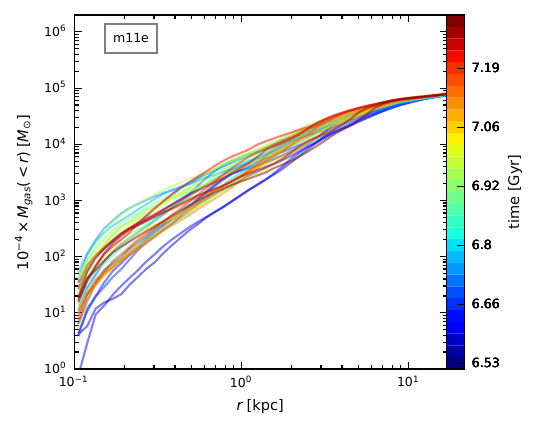}
    \includegraphics[width= 0.49\textwidth]{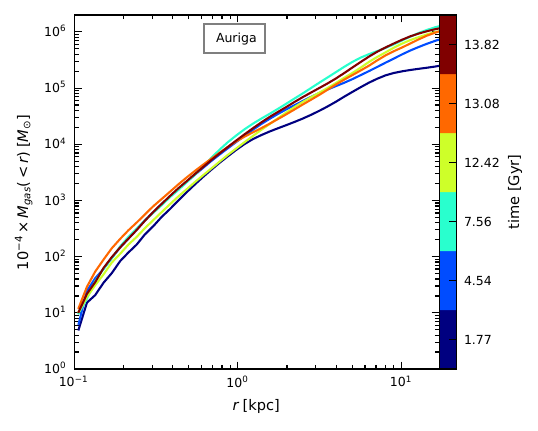} 
 \caption{Evolution of gas mass enclosed within radius $r$ for a FIRE-2 simulation halo (left), compared to halo~1 from Auriga (right). While the gas mass in the central regions of the FIRE halo undergoes large fluctuations, the Auriga halo shows only modest variations over a much longer period, driven primarily by accretion.}
\label{fig:massfluc}
\end{figure}
In Fig.~\ref{fig:denscont},  we compare the gas density contrasts in three haloes from the FIRE simulation with those in ISIM and halo~1 of the Auriga simulation. The lack of significant fluctuations in the latter relative to the FIRE and ISIM 
cases is clear. This can also be probed through measuring the temporal mass variation (Fig.~\ref{fig:massfluc}). 
Note that in the case of the Auriga halo we show the evolution over nearly a Hubble time, to illustrate that no significant evolution 
occurs other than due to mass accretion. 

\begin{figure}[H]
    \noindent\centering
    \includegraphics[width= 0.49\textwidth ]{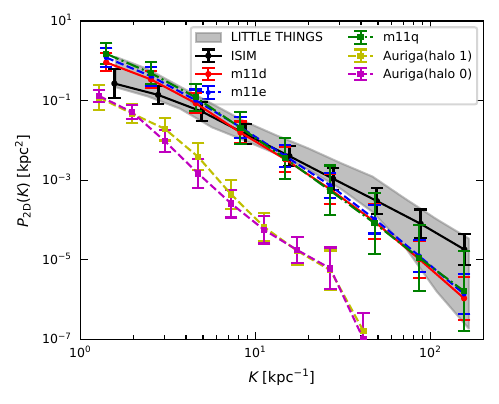}
    \includegraphics[width= 0.49\textwidth]{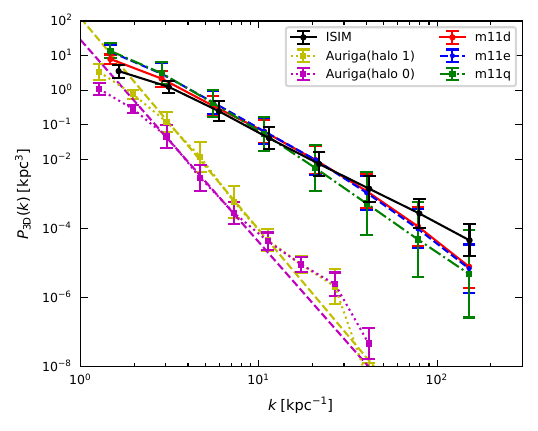}    
    \caption{Comparison of power spectra corresponding to the density contrasts shown in Fig.~\ref{fig:denscont}. We also include an additional halo from the Auriga simulation (level 3/low-mass dwarf, halo ID 0 in Table~A1 of~\cite{AURIGA2024}, referred to as halo~0 in the text) and the 2D galaxy spectra, 
    indicated by the shaded region.
    The box-size for ISIM is  $d =5$~kpc 
    while $d = 10~{\rm kpc}$ for FIRE and Auriga haloes, accounting for the larger effective scale of the gas distribution in these haloes, which are more massive than in ISIM (cf. Fig.~\ref{fig:2D3lin} in the case of the FIRE haloes). The dashed lines, showing power law fits to the limited 
    scaling region of the 3D Auriga spectra, have slopes  -6.1 for halo~1 and -5.8 for halo~0.}
\label{fig:cospower}
\end{figure}

Fig.~\ref{fig:cospower} compares the power spectra
associated with the density contrast of Fig.~\ref{fig:denscont}. 
For further verification of the qualitative 
difference between the gas spectra in Auriga compared to FIRE and ISIM, 
we also include an additional halo~0 from Auriga, which has smaller mass but also 
exhibited qualitatively the same density contrast as halo~1. 
All spectra  from  the  FIRE suite are much closer to those for ISIM (and to the galaxies) 
than in the Auriga simulation. The 3D spectra especially show the qualitative difference: the region where a power law may be identified is small, and its slope is extremely steep ($P_{3D} (k) \sim k^{-6}$). Indeed, the large scale power, which  is  an order magnitude 
 smaller than in FIRE, is largely due to the spiral patterns in a disc-like system rather than  the kind of scale-free energy cascade expected from  feedback-driven turbulence. 
A naive application of our model leads to  exceedingly large relaxation times. For halo~1, we may take 
$n = 6$, $v_c = 100~ {\rm km/s}$, 
$r_0 = 6.4~{\rm kpc}$, $\rho_0 (r_0) = 2 \times 10^6~M_\odot {\rm kpc^{-3}}$. This gives $t_{\rm relax} \simeq 1.5 \times 10^{13}$ years at $r_0$, 
or about 1100 Hubble times.   For the smaller 
halo~0---with  $P(k_m) = 0.1~{\rm kpc}^{-1},
$n = 6$, v_c = 35~ {\rm km/s}, 
r_0 = 5~{\rm kpc}, \rho_0 (r_0) = 5 \times 10^5~M_\odot {\rm kpc^{-3}}$--- one obtains $t_{\rm relax} \simeq 2 \times 10^{13}$ years at $r_0$. 
But the model does not  even strictly apply, 
due to the absence of a wavenumber regime 
$k_m \ll  k \ll k_x$ through which a scale-free cascade of fluctuations 
can be defined. 


\begin{figure}[H]
    \includegraphics[width=\textwidth]{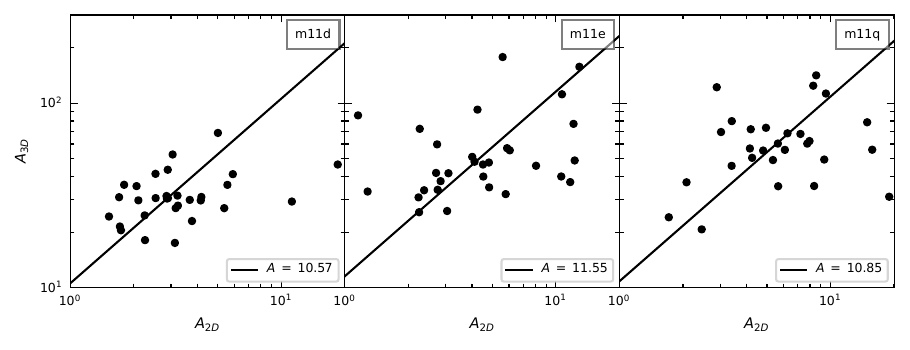}
    \includegraphics[width=\textwidth]{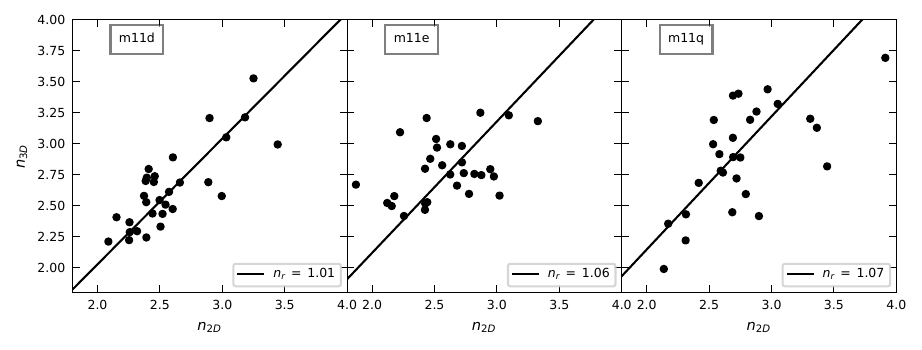}
    \caption{Relations between amplitudes and indices of power spectra 
    of gas fluctuations in the FIRE-2 simulation haloes, when fitted using equations~\ref{eq:PPS}.}
\label{fig:FIREA2DA3D}
\end{figure}

The relation between the 2D  and  3D spectra for the FIRE
haloes are shown in Fig.~\ref{fig:FIREA2DA3D}. These are analogous to the ones shown in Fig.~\ref{fig:PScalib} for ISIM, 
and also complementary, since the haloes in the FIRE simulation 
have larger effective gas radius. 
They will thus be used to verify our projection  model described below. 

\subsubsection{ Interpretation  in terms of projection model}
\label{sec:project}

We consider the following projection~\cite{Peacock1999coph.book.....P}
\begin{equation}
    P_{2D} (K)  = \frac{1}{\pi} \int_{0}^{\infty} P_{3D} (k) \tilde{W}^2 (k_z, L)
    d k_z,
\label{eq:project}
\end{equation}
where $k_{z}$ denotes the component of ${\bf k}$  that is vertical 
to the plane of projection (containing the radius vector ${\bf R}$), and $\tilde{W}$ 
refers to a filter corresponding to a slab of thickness 
$2 L$ (corresponding to the effective galaxy thickness) along that direction. 
We use a Gaussian 
filter, which  averages the fluctuations such that
$\delta_{2D} =  \frac{1}{\sqrt{\pi} L} \int \delta_{3D} e^{-z^2/2 L^2} dz$. 
The corresponding filter in Fourier space is  $\tilde{W}^2 = e^{-k_z^2 L^2}$. 
Note that $\delta_{2D}$, thus defined,  
corresponds
to our definition of $\delta_{2D} = \delta \Sigma ({\bf R})/\Sigma (R)$
when the average density $\rho (r)$ 
is constant over scales $\sim L$
(cf. first paragraph of Section~\ref{sec:ISIMcalib} for definition).
Thus, 
equation~(\ref{eq:project})
is  an  estimate of the effect of projection that 
is particularly effective when most of the power comes from radii where the density is slowly varying. 

\begin{figure}
    \centering
    \includegraphics[width= 11 cm]{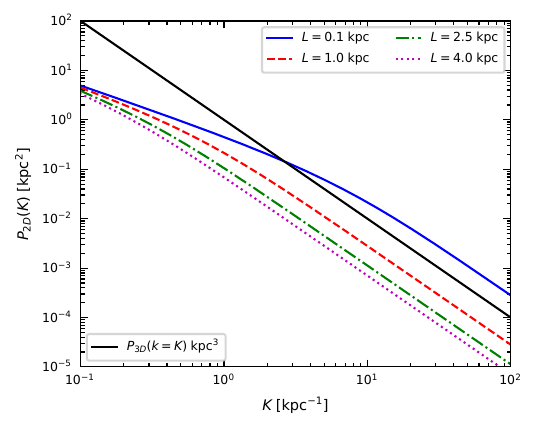}
    \caption{Conversion from 3D to 2D power spectrum using 
    equation~(\ref{eq:project}) for various values of the projection scale $L$.
   There are tow regimes: one
    with the a power index reduced by one, and another with $P_{3D} \sim L P_{2D}$. The latter being more 
    relevant to the simulations and sample of galaxies, as discussed in the text.} 
    \label{fig:2D3Deq}
\end{figure}

\begin{figure}
    \centering
    \includegraphics[width= 12 cm]{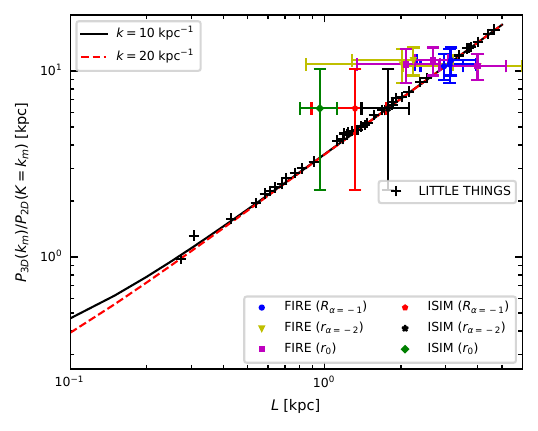}
    \caption{
Conversion between 2D and 3D power spectra. Crosses show the ratio of 3D to 2D spectra calculated using equation~(\ref{eq:project}) at \(k= k_m\) and \(L=r_0\) for the sample of galaxies. The values of \(r_{0}\) are as  on the x-axis of Fig.~\ref{fig:rho0}, and those of \(k_{m}\) are obtained as described in Section~\ref{sec:obsps}. The lines show the ratios of 3D and 2D spectra taken with \(k=K\) of 10 and 20~${\rm kpc}^{-1}$, which define the quasi-linear regime (where \(P_{\rm 3D} \sim L P_{\rm 2D}\)) for the range of \(L\) considered (this approximates the large $L$ limit 
in equation~\ref{eq:project}, which gives $P_{\rm 3D} = 2 \sqrt{\pi}L~P_{\rm 2D}$). The simulation results also lie in this regime; they are displayed as points with error bars, where the ratio of the spectra is calculated directly rather than via projection. They thus serve as a validation of the projection model. We show results referring to the different measures of the effective radial density scale discussed in Section~\ref{sec:rho0} (as summarized in Fig.~\ref{fig:collect_project_dens}). The values of \(r_{0}\) for the simulations are derived by measuring densities in cylinders containing \(90\%\) of the mass at \(R_{1/2}\) (effectively assuming $\eta = 0.9$ in equation~\ref{eq:r_o}, while \(q\) is automatically fixed from the 3D simulation data). The error bars associated with ISIM include the effect of variations in the box sizes over which the power spectra are calculated (Section~\ref{sec:ISIMcalib}).}  
    \label{fig:2D3lin}
\end{figure}

Fig.~\ref{fig:2D3Deq} shows the result of using 
equation~(\ref{eq:project}) for various values of 
$L$.  For illustration, the 3D power law has index $-2$ but the following trends are generic. 
Two regimes may be isolated. In the limit $L \ll 1/k$, the filter permits modes with $k_z \gg k$ to pass, rendering equation~(1) equivalent to an unfiltered integral over wavenumbers. In this case, a spectrum $P_{\mathrm{3D}} \propto k^{-n}$ is simply projected to $P_{\mathrm{2D}} \sim K^{-(n-1)}$ (this is the counterpart, in wavenumber space, to the surface density scaling as $1/R$ when the 3D density goes as $1/r^2$, mentioned in Section~\ref{sec:rho0}). In contrast, when the line-of-sight projection scale is large relative to the scale of interest ($L \gg 1/k$), the filter restricts $k_z$ to small values ($k_z \ll k$), yielding $P_{\mathrm{3D}} \sim L \, P_{\mathrm{2D}}$ (this approximates the  
large $L$ limit of equation~\ref{eq:project}, which is $P_{\rm 3D} = 2 \sqrt{\pi}L~P_{\rm 2D}$).    
In this quasi-linear limit, $n_{\mathrm{3D}} / n_{\mathrm{2D}}$ is near unity, matching the results of the ISIM and FIRE simulations discussed above.

We now verify that it is the quasi-linear regime
that is also relevant for converting the observed 2D power spectra of our sample of galaxies to 3D spectra.  
For this, we first need an estimate of the relevant $L$, 
which one 
may expect to be of the order of the effective scale 
of the density distribution, beyond which the gas 
density declines steeply.  
By inspecting Fig.~\ref{fig:2D3Deq}, we note 
that a ratio of the 3D to 2D power spectra 
of order $A \approx 6$--- as found from the calibration 
using the fiducial simulation ISIM 
(eq~\ref{eq:calib})--- 
corresponds to $L \gtrsim 1~{\rm kpc}$. 
This matches the energy transfer saturation 
scale $r_{\rm sat}$ and the 
closely related measures of the radial  
 scale setting the effective density, as  
 discussed in Section~\ref{sec:rho0}
(Fig.~\ref{fig:collect_project_dens}).  It  
suggests that the prediction of the simple projection
model matches  that of the simulation  when 
$L$ is indeed of the 
order of that effective scale.
As we will see below, the results of the 
analysis of the gas in the FIRE-2 haloes 
are also consistent with this.

For galaxies, the measure we adopted 
for the effective gas radial scale 
was $r_0$, which was derived from 
the half mass projected radius. 
We will now assume that the scaling 
$L = r_0$ holds for the galaxies and  
examine whether that renders them in the quasi-linear 
regime of conversion between 2D and 3D spectra. 
As noted above, 
significant departures from linearity in the conversion 
from 3D to 2D are clearest 
at smaller $k$. To verify their absence,  we thus calculate 
the ratio $P_{\rm 3D} (k_m)/P_{\rm 2D} (K = k_m)$ as a function of $L = r_0$ for each galaxy, 
using equation~(\ref{eq:project})
and the values of $k_m$ found in the subsection below. 
The results are shown in Fig.~\ref{fig:2D3lin}. 
There we  also plot 
two lines where equation~(\ref{eq:project}) is used with 
large values of $k_m$, 
defining the quasi-linear regime, as reference.
As smaller galaxies have smaller $L = r_0$
but larger $k_m$, 
the ratio between the 3D and 2D spectra invariably 
lies in the nearly linear 
segment of the relation ($r_0 \gg 1/k_m$). 
We thus infer that the galaxies lie in the same 
quasilinear projection regime as the gas 
in ISIM and also the FIRE haloes, which are seen to be consistent with the projection model within the error bars.

Given the general consistency of the 
projection model with the simulation results
(with $L = r_0$), and that the galaxies lie
in the same quasilinear regime $P_{3 D} \sim  L P_{2D}$
as the simulations, 
we replace equation~(\ref{eq:project}) 
by extrapolating the results of the calibration 
from the simulations as
\begin{equation}
    P_{\rm 2D} (K = k) =  
    A^{-1}~\beta~
    ~P_{\rm 3D} (k)~ \left(\frac{k}{{\rm kpc}^{-1}}\right)^{(n_{3D} - n_{2D})} {\rm kpc}^{-1}, 
\label{eq:convert}
\end{equation}
where $P_{\rm 2D}$ and $P_{\rm 3D}$ are assumed to be given as in equations (\ref{eq:PPS}).
This is readily inverted and solved for $P_{\rm 3D}$, 
given $P_{\rm 2D}$.   
The mean values and errors on $P_{\rm 2D}$ (i.e on the normalization $A_{\rm 2D}$ and index $n_{\rm 2D}$) 
are obtained as described in Section~\ref{sec:obsps} and Section~\ref{sec:errors} 
(item~2). 
The factor $\beta = r_{\rm sat}/r_0$ scales 
the estimated effective density scale 
of each galaxy relative to the 
effective density scale 
in ISIM. 
The means and errors
of $r_0$ for each galaxy 
are obtained as described in 
Section~\ref{sec:rho0} 
and Section~\ref{sec:errors}. 
The weak 
departure from linearity is incorporated through
the index $n_{\rm 3D} - n_{\rm 2D} = n_{\rm 2D} (n_r - 1)$. 
As we have not comprehensively 
investigated the results of the FIRE simulations, 
compared to ISIM, we only use the latter
for estimating $\beta$, as well $A$ and $n_r$ (as given by equations 
\ref{eq:calib}).

\subsection{Observed spectra and minimal wavenumber}
\label{sec:obsps}

\begin{figure}
    \centering
    \includegraphics[width=\linewidth]{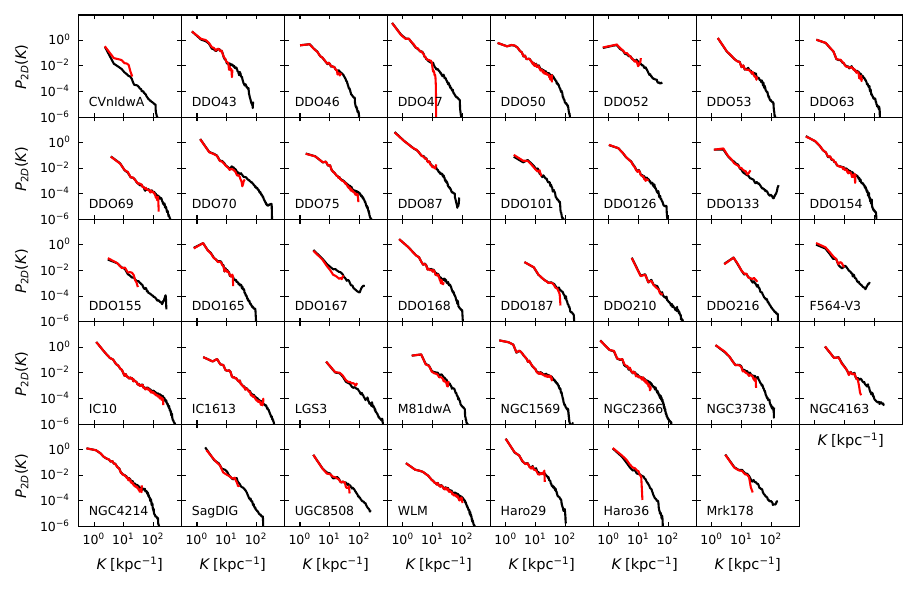}
    \caption{Comparison between power spectra 
    of the relative gas surface density fluctuations in the LITTLE THINGS  galaxy sample at different pixel sizes: at the observational resolution 
    limit, corresponding to the beam size  (in red); and at the reconstructed image pixel resolution (in blue).
    The spectra are evaluated inside boxes corresponding to radii containing 
    $90~\%$
    of the projected gas mass.}
    \label{fig:Pixel_obs}
\end{figure}

\begin{figure}
    \centering
    \includegraphics[width=\linewidth]{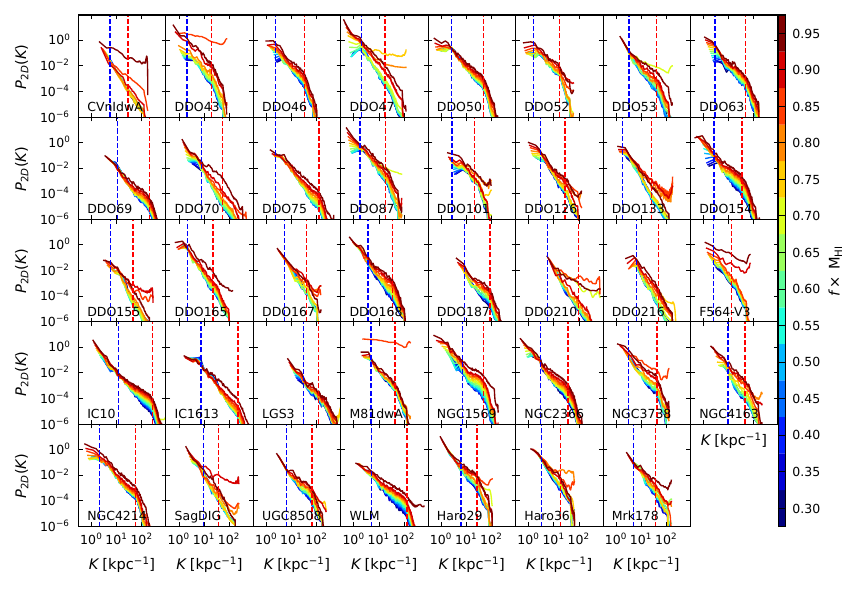}
    \caption{Convergence of power spectra of  relative surface density 
    fluctuations. The lines correspond to spectra taken over boxes with sides twice the radii enclosing a specified fraction, \(f\), of the total observed gas mass, \(M_{\mathrm{H}I}\). The vertical blue lines at small \(K\) refer to the maximal fluctuation scales, as inferred from the H~I cloud sizes given in ref.~\cite{2019AJ....157..241H}. The vertical red lines at large \(K\) refer to the observational beam resolution scale. Note that, due to the observational sensitivity limit, all of the observed gas mass (\(f=1\)) is generally enclosed in boxes corresponding to \(d \lesssim 6~{\rm kpc}\) when scaled to the fiducial simulation ISIM (cf. Fig.~\ref{fig:psbox} and the discussion in the text).}
    \label{fig:BOXPS_obs}
\end{figure} 

To produce power spectra from the galaxy sample data, 
we  use the image maps (cf. Section~\ref{sec:sudens}).  
The use of the image maps, 
rather than the interferometric data  directly~\cite{2009MNRAS.398..887D},
is preferred
because we are interested 
in quantifying  relative fluctuations of the surface density 
$\delta 
\Sigma/\Sigma = \Sigma({\bf R})/\Sigma(R) - 1$, and therefore 
positional information in real space is 
required (in order to obtain a density field $\Sigma(\bf{R})$ and scale it  
with $\Sigma (R)$ before taking the spectrum). Discussion and 
comparison of the two methods for some of the galaxies studied here can be found in ref~\cite{2012ApJLTPS}.

As with the simulation ISIM, we checked that varying  the  
pixel size over which the power spectra are evaluated
does not significantly affect the power spectra. 
Fig.~\ref{fig:Pixel_obs} compares the beam scale power spectra 
to the cleaned image pixel scale spectra. 
The fluctuations at the largest scales, which are crucial to our 
estimates of the relaxation time, again remain relatively stable and robust. 
 On the other hand, small scale noise is smoothed and attenuated at large 
$K$ by the Multiscale CLEAN algorithm used to produce the images (Section~\ref{sec:errors}). 

We also varied the box size and 
studied the convergence of the spectra. 
As with the simulation, when fixing the box size a compromise is made.  
 The box size  $d$ has to be large enough
to capture the largest fluctuations at minimal wavenumber $k_m$. For the observations, the  outer limit is
effectively set by requiring $d$ not to be so large as to 
introduce spurious discreteness effects and noise  near the observational minimal surface density limit 
of $10^{19}$ atoms per ${\rm cm^{2}}$. 
For our sample, this generally 
corresponds to radii $\lesssim 2 R_{1/2}$, 
which for ISIM
(where $R_{1/2} \simeq 1.5 {\rm kpc}$),
 translates to $d \lesssim 6 {\rm kpc}$. For the galaxies we are therefore invariably within the box sizes considered 
in Fig.~\ref{fig:psbox}, even when the FFT is taken over the whole observed gas distribution.

Fig~\ref{fig:BOXPS_obs} shows the convergence of the 
spectra for various radii comprising different percentages of the enclosed projected gas mass.  We find general
convergence when $d/2 \gtrsim R_{90}$, where
$R_{90}$ is the cylindrical radius containing $90 \%$ of the observed gas mass. 
 To test this, and obtain an error estimate, 
we estimated the mean and variance of the distribution of 
the power spectra fitting parameters for half box sizes between 
radii containing $85 \%$ and $95 \%$  of the mass, as described in Section~\ref{sec:errors}. 
The slopes and normalization of the power law parts of the galaxy spectra 
are determined starting at the midpoint between 
the two vertical lines shown in Fig.~\ref{fig:BOXPS_obs} 
(i,e, between the minimal wavenumber 
and that corresponding to the beam scale), and applying  the same 
procedure described above in relation to the 
simulations (equation~14).

In order to complete the quantification
of the ingredients in equation (\ref{eq:relax}), 
we still need to estimate the wavenumber associated with the largest fluctuation scale 
$k_m$.  We do not have direct access to the 3D power spectra of 
the observed galaxies, and the turnover from the power law at large scales
in the 2D spectrum can, in principle, result from projection. 
 Moreover, power at small wavenumbers 
 can arise from large scale asymmetries and spiral structures, rather than 
 the scale-free cascade of fluctuations relevant to our model, 
 as was apparent in the case of the Auriga simulation gas.
It may therefore be more accurate to attempt  
to infer $k_m$ in equation~\ref{eq:relax}
in an independent manner. 
To this end, we use the results of ref.~\cite{2019AJ....157..241H}, 
which determined maximal cloud sizes for galaxies in our sample. 
This has the added benefit that the largest sizes more likely correspond
to actual hydrodynamic fluctuations rather than from structures that may
arise from large scale deformations, which may not be 
as relevant to our model. In this way, we associate the radius $R_{\rm cloud}$ with $1/k_m$
(noting that the cloud radius corresponds to 3D fluctuations 
characterized by wavenumbers $k$, rather than
projected fluctuations expressed in terms of $K$). We also note that, for the purpose of illustration, 
in Fig.~\ref{fig:BOXPS_obs} we 
transform  the inferred $k_m$ into $K_m$
assuming isotropic fluctuations 
($\langle K^2 \rangle = \langle \frac{2}{3} k^2 \rangle$), such that the vertical lines correspond to  $K_m = \sqrt{2/3} k_m$. But this transformation does not enter into the calculation of the relaxation time using equation~(\ref{eq:relax}), as this depends directly on the 3D spectrum (and thus involves an inverse transformation from $K$ to $k$). The relation between $P_{3D} (k_m)$ and $P_{2D} (K = k_m)$ over the power law part of the spectra is thus 
obtained  using equation (\ref{eq:convert}).

\subsection{Error estimation and propagation}
\label{sec:errors}

\begin{figure}[H]
    \includegraphics[width= 0.49\textwidth]{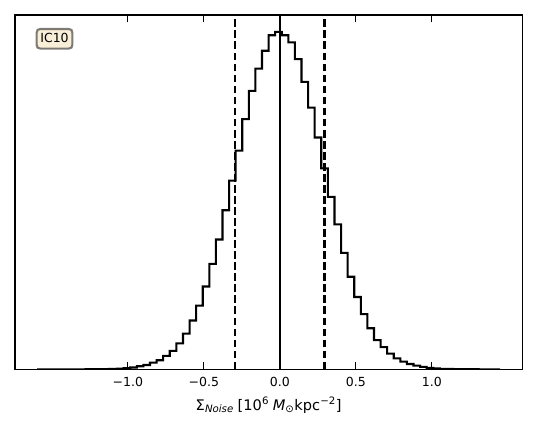}
    \includegraphics[width= 0.49\textwidth]{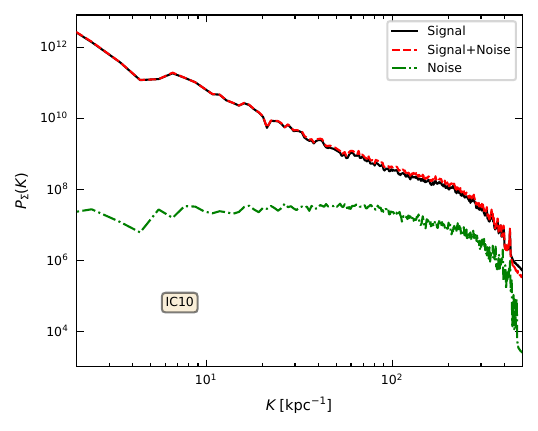}
    \caption{Distribution of the surface density residual noise (left) in a signal-free channel of the dwarf galaxy IC10, as an example, and the corresponding noise power spectrum relative to that of the signal (right). The noise is essentially white, attenuated on sub-beam scales by the Multiscale CLEAN algorithm.}
\label{fig:Noise}
\end{figure}

Equation~(\ref{eq:relax}) contains two terms related to the 
power spectra, namely $P(k_m)$ and $n$;  
another two related to the mass model ($v_p$ and $v_r$); 
in addition to the effective gas density $\rho_0$.
We  calculated error estimates  
on all these quantities, some of which were mentioned above when relevant. 
All errors were propagated through a Monte Carlo procedure. We adopt the Latin-hypercube MC 
sampling and perform non-order-specific error propagation via the Python package mcerp~\footnote{https://pypi.org/project/mcerp/}. The MC sampling is conducted  over 
a number of points of order $10^6$.

We note that we do not place error estimates on the HI images themselves, as these are obtained from 
the 'dirty' beams using Multiscale CLEAN \citep{2012AJ....144..134H}, which is an inherently 
nonlinear procedure with no readily quantifiable errors. A reassuring  result is that the final 
residuals are small and 
consist of nearly perfectly Gaussian white noise. The residual are furthermore subtracted in the 
noise-free images we use here, but adding them gives an idea of the errors involved. We estimate these 
from the robust data-cube signal-free channels (as described in CASA-imstat 
task~\citep{2022PASP..134k4501C}). A typical example is shown in Fig.~\ref{fig:Noise} where it is clear that the effect is minimal.  

The error estimates that were included in the calculations are as follows.  

\begin{enumerate}

\item 
The errors on the mass model are obtained from 
ref.~\cite{2015AJ....149..180O},
where 
Table~2 gives the core radius and characteristic density 
of the isothermal halo fits to the data. Monte Carlo distributions
are produced for these quantities 
and the circular velocity at the core 
radius $v_c$
and $v_r = d~\sqrt{G\bar{\rho}_{\rm DM}(d/2)}/2$ are estimated by propagating the errors 
as described above. 

\item 

Errors on the observed (2D) power spectra include those on the normalization $A_{2D}$ and the index $n_{2D}$ of the power law fits. 
We also include in our estimate the error entering from 
our assumed convergence box size, which has sides 
 $\simeq 2 \times R_{90}$ (where $R_{90}$
is the radius containing 90 percent of the projected mass). 
To this end, we take a distribution of power spectra including  
radii between $R_{85}$ and $R_{95}$, fit  power laws to 
this sample of spectra and calculate the mean values---which are 
statistically convergent and usually centred close to the value at $R_{90}$---as well as 
the variances of the normalization and index of the spectrum. 

\item 

We use the isolated dwarf galaxy 
simulation (ISIM) to estimate the 
error in the conversion of the 2D spectra into 3D. 
Similar to the above procedure, we stack all power spectra from all snapshots and including all box sizes used (i.e., 3, 4, 5, and 6~kpc), and estimate the errors in the conversion fits to the normalization and index conversion factors $A_{3D}/A_{2D}$ and $n_r = n_{3D}/n_{2D}$ (Fig.~\ref{fig:PScalib}).
This leads to equations~(\ref{eq:calib}).

\item

We then invert equation~(\ref{eq:project}) to obtain the 3D power spectra, 
including the errors in $n_{2D}$, calibration errors in 
$A$ and  $n_r$, and also errors in $r_0$ (described below). 
This is done through equation~(\ref{eq:convert}), which is verified by the 
data and simulation points (including from FIRE-2) in Fig.~\ref{fig:2D3lin}. 

We finally introduce mean values and errors in $k_m$ from the maximum cloud size values, as given 
in~\cite{2019AJ....157..241H}. These are propagated  to obtain estimates of 
$P(k_m)$. 

\item

The effective radius $r_0$ and density $\rho_0 (r_0)$, as given by equation~(\ref{eq:rho0}), incorporate errors
in the flattening $q$, which for each individual galaxy is obtained from ref.~\citep{2017MNRAS.465L..49J}. 
It moreover includes the uncertainty in the density distribution, parametrized by $\eta$, 
for which we use a flat prior comprising reasonably expected values, $0.71 \le \eta \le  0.95$, as discussed in Section~\ref{sec:rho0}. 

\item 
For Fig.~\ref{fig:relaxM200}, the virial masses are estimated as in~\cite{2015AJ....149..180O}, by assuming an initial NFW~\cite{NFW} profile and associating its \(V_{200}\) with the maximal speed of the isothermal halo fit, then using the empirical relation \(M_{200} \simeq 3.29 \times 10^5 V_{200}^3\)~\cite{calib200centralslopes2011AJ}. The errors on \(M_{200}\) are propagated from those on \(V_{200}\) (as given in Table 2 of~\cite{2015AJ....149..180O}). 

\end{enumerate} 

Table~1 displays the mean values 
with uncertainties of quantities entering into 
equation~(\ref{eq:relax}).

\begin{table}[h]
\centering
\renewcommand{\arraystretch}{1.5}
\resizebox{\textwidth}{!}{%
\begin{tabular}{|l|c|c|c|c|c|c|c|c|c|}
\toprule
                                                         Galaxy & $r_{0}$ & log($\rho_{0})$ & $n^{\rm Obs}$ & $n^{\rm Obs+Calib}$ & $P^{\rm Obs}_{3D}(k_{m})$ & $P^{\rm Obs+Calib}_{3D}(k_{m})$ & $v_p$ & $v_r$ & log($t_{\rm relax}$) \\
\midrule
   CVnIdwA & $0.73 \pm 0.15$ & $7.25_{-0.17}^{+0.25}$ & $3.03 \pm 0.28$& $3.03 \pm 0.37$ & $0.04_{-0.02}^{+0.03}$ & $0.04_{-0.02}^{+0.05}$ & $19.77_{-5.35}^{+5.68}$ & $8.06_{-1.08}^{+0.95}$ & $0.09_{-0.76}^{+0.64}$ \\
DDO43 & $2.14 \pm 0.45$ & $6.41_{-0.17}^{+0.25}$ & $2.49 \pm 0.23$& $2.49 \pm 0.30$ & $14.10_{-5.62}^{+9.06}$ & $14.07_{-7.11}^{+14.18}$ & $18.54_{-4.26}^{+4.57}$ & $16.96_{-3.61}^{+3.73}$ & $-0.59_{-0.71}^{+0.58}$ \\
 DDO47 & $5.02 \pm 0.89$ & $5.67_{-0.15}^{+0.22}$ & $2.68 \pm 0.12$& $2.68 \pm 0.24$ & $17.79_{-4.31}^{+5.42}$ & $17.73_{-7.61}^{+13.20}$ & $44.95_{-6.30}^{+6.54}$ & $33.14_{-3.49}^{+3.40}$ & $1.90_{-0.55}^{+0.45}$ \\
     DDO50 & $3.12 \pm 1.03$ & $6.46_{-0.22}^{+0.33}$ & $1.96 \pm 0.07$& $1.96 \pm 0.16$ & $2.94_{-0.59}^{+0.73}$ & $2.94_{-1.22}^{+2.09}$ & $8.34_{-8.33}^{+8.69}$ & $8.66_{-8.64}^{+8.88}$ & $-1.39_{-0.00}^{+1.27}$ \\
    DDO52 & $3.99 \pm 0.16$ & $5.85_{-0.05}^{+0.05}$ & $1.98 \pm 0.42$& $1.98 \pm 0.44$ & $9.94_{-3.77}^{+5.99}$ & $9.93_{-4.87}^{+9.54}$ & $32.32_{-2.05}^{+2.10}$ & $28.18_{-1.58}^{+1.59}$ & $1.35_{-0.35}^{+0.35}$ \\
  DDO53 & $1.10 \pm 0.23$ & $6.76_{-0.17}^{+0.25}$ & $2.07 \pm 0.35$& $2.07 \pm 0.38$ & $0.71_{-0.31}^{+0.53}$ & $0.71_{-0.38}^{+0.80}$ & $38.88_{-26.19}^{+33.44}$ & $18.64_{-8.43}^{+4.31}$ & $0.51_{-1.47}^{+0.91}$ \\
    DDO70 & $1.32 \pm 0.28$ & $6.42_{-0.17}^{+0.25}$ & $2.37 \pm 0.18$& $2.37 \pm 0.25$ & $0.10_{-0.04}^{+0.08}$ & $0.10_{-0.05}^{+0.12}$ & $19.03_{-4.89}^{+5.29}$ & $17.12_{-4.04}^{+4.16}$ & $1.55_{-0.77}^{+0.64}$ \\
 DDO87 & $3.96 \pm 0.84$ & $5.77_{-0.17}^{+0.25}$ & $2.42 \pm 0.14$& $2.42 \pm 0.23$ & $18.17_{-5.46}^{+7.45}$ & $18.10_{-8.15}^{+14.66}$ & $31.93_{-1.66}^{+1.69}$ & $25.16_{-1.05}^{+1.05}$ & $1.25_{-0.59}^{+0.45}$ \\
   DDO126 & $1.84 \pm 0.07$ & $6.58_{-0.05}^{+0.05}$ & $2.78 \pm 0.46$& $2.78 \pm 0.51$ & $2.11_{-0.95}^{+1.66}$ & $2.11_{-1.15}^{+2.45}$ & $21.47_{-1.87}^{+1.92}$ & $16.68_{-1.17}^{+1.16}$ & $0.13_{-0.42}^{+0.42}$ \\
   DDO133 & $1.72 \pm 0.13$ & $6.53_{-0.09}^{+0.10}$ & $3.08 \pm 0.56$& $3.08 \pm 0.60$ & $4.54_{-1.79}^{+2.94}$ & $4.54_{-2.28}^{+4.56}$ & $24.75_{-2.16}^{+2.23}$ & $20.88_{-1.58}^{+1.59}$ & $0.15_{-0.41}^{+0.40}$ \\
   DDO154 & $3.14 \pm 0.13$ & $6.19_{-0.06}^{+0.05}$ & $1.84 \pm 0.27$& $1.84 \pm 0.31$ & $2.41_{-1.09}^{+1.99}$ & $2.41_{-1.30}^{+2.82}$ & $24.12_{-1.04}^{+1.05}$ & $22.87_{-0.93}^{+0.94}$ & $0.91_{-0.39}^{+0.38}$ \\
 DDO168 & $1.80 \pm 0.38$ & $6.85_{-0.17}^{+0.25}$ & $2.15 \pm 0.09$& $2.15 \pm 0.19$ & $0.55_{-0.15}^{+0.20}$ & $0.55_{-0.25}^{+0.44}$ & $61.18_{-18.37}^{+19.13}$ & $40.13_{-8.11}^{+6.74}$ & $1.27_{-0.74}^{+0.58}$ \\
    DDO210 & $0.28 \pm 0.06$ & $7.15_{-0.17}^{+0.25}$ & $2.71 \pm 0.58$& $2.71 \pm 0.62$ & $0.11_{-0.07}^{+0.18}$ & $0.11_{-0.08}^{+0.23}$ & $7.42_{-1.95}^{+2.07}$ & $6.02_{-1.31}^{+1.26}$ & $-1.22_{-0.89}^{+0.78}$ \\
    DDO216 & $0.42 \pm 0.07$ & $7.08_{-0.14}^{+0.20}$ & $2.81 \pm 0.87$& $2.81 \pm 0.89$ & $0.21_{-0.17}^{+0.80}$ & $0.21_{-0.18}^{+0.92}$ & $5.74_{-1.52}^{+1.63}$ & $5.55_{-1.41}^{+1.47}$ & $-1.60_{-1.06}^{+1.00}$ \\
  F564-V3 & $1.73 \pm 0.08$ & $6.12_{-0.06}^{+0.06}$ & $1.80 \pm 0.63$& $1.80 \pm 0.65$ & $2.69_{-1.72}^{+4.37}$ & $2.69_{-1.82}^{+5.23}$ & $15.36_{-7.44}^{+8.20}$ & $13.96_{-6.44}^{+6.52}$ & $0.31_{-1.05}^{+0.84}$ \\
     IC10 & $1.29 \pm 0.27$ & $6.61_{-0.17}^{+0.25}$ & $1.50 \pm 0.19$& $1.50 \pm 0.22$ & $0.03_{-0.01}^{+0.02}$ & $0.03_{-0.02}^{+0.03}$ & $12.28_{-5.99}^{+6.77}$ & $12.06_{-5.70}^{+6.15}$ & $0.85_{-1.16}^{+0.83}$ \\
    IC1613 & $1.47 \pm 0.04$ & $6.25_{-0.05}^{+0.04}$ & $2.35 \pm 0.22$& $2.35 \pm 0.28$ & $0.14_{-0.05}^{+0.08}$ & $0.14_{-0.07}^{+0.15}$ & $3.03_{-0.65}^{+0.68}$ & $2.97_{-0.60}^{+0.62}$ & $-0.56_{-0.47}^{+0.46}$ \\
NGC1569 & $1.82 \pm 0.05$ & $6.84_{-0.04}^{+0.04}$ & $1.65 \pm 0.15$& $1.65 \pm 0.20$ & $0.33_{-0.11}^{+0.16}$ & $0.33_{-0.16}^{+0.31}$ & $36.49_{-11.09}^{+11.52}$ & $26.43_{-5.94}^{+5.09}$ & $0.84_{-0.53}^{+0.46}$ \\
  NGC2366 & $2.84 \pm 0.26$ & $6.70_{-0.10}^{+0.13}$ & $1.69 \pm 0.21$& $1.69 \pm 0.25$ & $0.73_{-0.24}^{+0.36}$ & $0.73_{-0.35}^{+0.65}$ & $27.90_{-1.21}^{+1.22}$ & $25.64_{-1.03}^{+1.03}$ & $0.54_{-0.40}^{+0.38}$ \\
  UGC8508 & $0.62 \pm 0.02$ & $7.13_{-0.04}^{+0.04}$ & $2.47 \pm 0.55$& $2.47 \pm 0.58$ & $0.05_{-0.04}^{+0.11}$ & $0.05_{-0.04}^{+0.13}$ & $45.66_{-5.01}^{+5.06}$ & $16.38_{-0.57}^{+0.54}$ & $1.23_{-0.64}^{+0.64}$ \\
      WLM & $1.10 \pm 0.05$ & $6.94_{-0.06}^{+0.06}$ & $1.56 \pm 0.04$& $1.56 \pm 0.13$ & $0.06_{-0.01}^{+0.02}$ & $0.06_{-0.03}^{+0.05}$ & $19.53_{-0.37}^{+0.37}$ & $16.67_{-0.29}^{+0.29}$ & $0.60_{-0.28}^{+0.28}$ \\
   Haro29 & $1.55 \pm 0.33$ & $6.40_{-0.17}^{+0.25}$ & $1.69 \pm 0.37$& $1.69 \pm 0.39$ & $0.29_{-0.16}^{+0.33}$ & $0.29_{-0.18}^{+0.43}$ & $15.97_{-3.86}^{+4.07}$ & $16.04_{-3.81}^{+3.96}$ & $0.77_{-0.82}^{+0.71}$ \\
\bottomrule
\end{tabular}

}
\caption{Means and errors of the quantities used to evaluate the relaxation time from equation~(\ref{eq:relax}). For the spectral index $n$ and normalization $P(k_m)$, the superscript ''Obs'' refers to the inclusion of errors from the 2D spectra only, while columns labelled ''Obs + Calib'' also include calibration errors. These are introduced through equation~(\ref{eq:calib}) and propagated as described in the text. The relaxation times quoted in the last column on the right include both.}
\label{tab:summary}
\end{table}

\subsection{The inner halo density slopes}
\label{sec:slopes}

\begin{figure}[H]
    \centering
    \includegraphics[width= 13 cm]{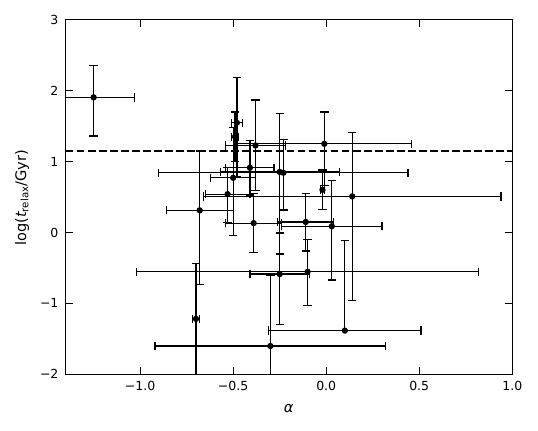}
\caption{
    The inner dark matter slopes of  haloes of the galaxies considered here, as found in~\cite{2015AJ....149..180O}, against the relaxation times calculated. 
Except for DDO~47, which has a relatively steep inner cusp and large relaxation time, and DDO~168, 
which has significant inner baryon fraction and slope $\alpha \simeq +1$ 
(and is omitted here), the points lie within  
a vertical column with largely 
overlapping error bars and with central values averaging to  
$\alpha =  -0.28$.}
    \label{fig:slopes}
\end{figure} 

In the model used here 
the relaxation time goes to zero as one approaches 
the centre of an initial NFW~\cite{NFW} halo, where both circular speeds and velocity dispersions (and hence $v_p$) go to zero. Whenever our mechanism is effective, a small core should therefore immediately form, expanding outwards with time as in Fig.~\ref{fig:core_relax} (although its  
size is eventually limited by the energy transfer 
radius $r_{\rm sat}$, and also the NFW scale length $r_s$; cf.
ref.~\cite{HEZFRC} Appendix~C).

Thus the expectation 
is to find nearly constant density cores, unless 
the relaxation time is exceedingly large 
(or recent core reformation occurred),
so as to make a core undetectable at any resolvable radius.  

In ref.~\cite{2015AJ....149..180O}, both the cored isothermal mass models and  NFW halo profiles were tried in order to fit the data. It was found that, except for a few cases, the latter were entirely unsuitable (e.g., resulting in negative concentration parameters $c$). 
The inner density profiles obtained do vary significantly because of the observational errors, but the median profiles 
were indeed consistent with an inner nearly constant density core (Fig.~5 in \cite{2015AJ....149..180O}). 
An attempt was also made to estimate the inner logarithmic slopes of the inner density profiles. 
This is necessarily sensitive to observational errors, as measuring the logarithmic slope effectively requires evaluating a second derivative of the rotation curve (and the estimation of the inner radius and its associated error involves a third derivative).

We plot the result of their analysis against the relaxation time in Fig.~\ref{fig:slopes}. Except for DDO~47, the points cluster in a column, 
with larger error bars and central values of 
$-0.28$. Many are consistent with 
a flat core. 
Some of the more negative slopes can in principle 
result from resolution effects, 
as the transition from
$\alpha = -2$ to $\alpha = 0$ 
occurs over more than an order of magnitude in radius~\cite{deblokmcgbosmarubin2001, dblokbosma2002A&A, 2015AJ....149..180O}.  Except for DD~47, the relaxation times are of the order of Hubble time or less. 

The case of DDO~47 is interesting because the rotation 
curve from  which the slope is inferred actually exhibits 
a nearly flat inner part, corresponding to a steep 
density cusp. This is followed, at larger radii, by
an intermediate region where the 
inferred circular velocity rises in accordance with 
a nearly constant density distribution, before flattening again (cf. Fig. 19 in \cite{2015AJ....149..180O}). In principle, the 
associated apparent flattening of the density distribution at intermediate radii
could be due to non-circular motion, which may  be in turn induced by a triaxial halo, a warp or a spiral 
structure~\cite{LittleThingsnoncirc2019AJ}. The complex behaviour of the rotation 
curve of DDO~47 is not apparent in earlier work, 
where such complications were found to be minor and the  curve well fitted by a model with a large core (of 7 kpc~\cite{burkertsalDDO472005ApJ}). This core (which is larger still compared to 
the mean value $r_c = 3.72 {\rm kpc}$ 
used here)  cannot be formed 
through the mechanism discussed in the present work, unless the fluctuation levels or the gas density were far larger in the past. If, on the other hand, the strong inner cusp with $\alpha \simeq -1.25$ is the more faithful description of the inner profile, this would be in accordance with the model presented here. 
For, using column 2 and 6 of Table 2 of~\cite{2015AJ....149..180O}, 
one derives a resolution
of $0.26 {\rm kpc}$ for this galaxy. On the other hand, to bring down the 
relaxation time to within a  
Hubble time we need to decrease $v_c^2$ in equation~(\ref{eq:relax}) 
by about 14 times. From Fig.~19 of ref.~\cite{2015AJ....149..180O}, 
we  find that this brings the core inside the resolution limit, and therefore cannot be detected. 



\backmatter


\bmhead{Data availability}
The observational HI data cubes and moment maps analyzed in this study are publicly available from the NRAO LITTLE THINGS repository (\url{https://science.nrao.edu/science/surveys/littlethings/data/data}). The cosmological simulation datasets are accessible through the public data releases of the FIRE-2 project (\url{https://fire.northwestern.edu/data/}) and the Auriga Project (\url{https://wwwmpa.mpa-garching.mpg.de/auriga/data.html}). All other intermediate data supporting the plots and conclusions of this study are available from the corresponding author upon reasonable request.

\bmhead{Code availability}
A Python package to calculate the power spectra of observed and simulated galaxies is publicly available on GitHub at \href{https://github.com/mwhashim/GalaxyPower}{GalaxyPower}.

\bibliography{gasdensfluct_nl}

\bmhead{Correspondence}
Correspondence and requests for materials should be addressed to A.~A.~El-Zant (email: \href{mailto:amr.elzant@bue.edu.eg}{amr.elzant@bue.edu.eg}).


\bmhead{Author contributions}
MH undertook all analysis of simulation and observational data and wrote Section 2.1 of the Methods. AAEZ. conceived the project and led the interpretation of the results, in collaboration with M.H. and FC, and wrote the bulk of the paper. F.C. guided the analysis and interpretation of the observational data and contributed to the text and its finalization. 

\bmhead{Funding}
This work was supported by the Science and Technology and Innovation Funding Authority (STDF), Egypt, under grant number 48289.

\bmhead{Competing interests}
The authors declare no competing interests.

\end{document}